**Strong magnon-magnon coupling in an ultralow damping all-magnetic-insulator heterostructure**

Jiacheng Liu*[1,5], Yuzan Xiong*[2], Jingming Liang*[3], Xuezhao Wu[1,5], Chen Liu[4], Shun Kong Cheung[1,5] Zheyu Ren[1,5], Ruizi Liu[1,5], Andrew Christy[2], Zehan Chen[1,5,6], Ferris Prima Nugraha[1,5], Xi-Xiang Zhang[4], Chi Wah Leung[3], Wei Zhang[#2], Qiming Shao[#1,5,6]

[1]Department of Electronic and Computer Engineering, The Hong Kong University of Science and Technology, Hong Kong SAR

[2]Department of Physics and Astronomy, The University of North Carolina at Chapel Hill, Chapel Hill, NC 27599, USA

[3]Department of Applied Physics, The Hong Kong Polytechnic University, Hong Kong SAR

[4] Physical Science and Engineering Division (PSE), King Abdullah University of Science and Technology (KAUST), Thuwal 23955–6900, Saudi Arabia

[5]IAS Center for Quantum Technologies, The Hong Kong University of Science and Technology, Hong Kong SAR

[6]Department of Physics, The Hong Kong University of Science and Technology, Hong Kong SAR

* Equal contributions # Corresponding emails: zhwei@unc.edu; eeqshao@ust.hk

**Magnetic insulators such as yttrium iron garnets (YIGs) are of paramount importance for spin-wave or magnonic devices as their ultralow damping enables ultralow power dissipation that is free of Joule heating, exotic magnon quantum state, and coherent coupling to other wave excitations. Magnetic insulator heterostructures bestow superior structural and magnetic properties and house immense design space thanks to the strong and engineerable exchange interaction between individual layers. To fully unleash their potential, realizing low damping and strong exchange coupling simultaneously is critical, which often requires high quality interface. Here, we show that such a demand is realized in an all-insulator thulium iron garnet (TmIG)/YIG bilayer system. The ultralow dissipation rates in both YIG and TmIG, along with their significant spin-spin interaction at the interface, enable strong and coherent magnon-magnon coupling with a benchmarking cooperativity value larger than the conventional ferromagnetic metal-based heterostructures. The coupling strength can be tuned by varying the magnetic insulator layer thickness and magnon modes, which is consistent with analytical calculations and micromagnetic simulations. Our results demonstrate TmIG/YIG as a novel platform for investigating hybrid magnonic phenomena and open opportunities in magnon devices comprising all-insulator heterostructures.**

Spin-wave (or magnonic) devices utilize magnon spin degree of freedom to process information, which can occur in magnetic insulators free from any charge current, and therefore, are promising contenders for ultralow-power functional circuits [1–4]. Magnetic garnets such as yttrium iron garnet ($Y_3Fe_5O_{12}$, YIG) have an ultralow damping factor, and they have enabled long magnon spin transmission [5], efficient magnon spin current generation [6], and magnon logic circuits [2,3]. Another type of magnetic garnet, thulium iron garnet ($Tm_3Fe_5O_{12}$, TmIG), has been engineered to a binary memory with a robust perpendicular magnetic anisotropy [7,8]. Besides, TmIG thin films can exhibit topological magnetic skyrmion phase [9,10], promising for future magnetic insulator-based racetrack memory devices. In addition to these promising practical applications, magnetic insulators are well-known for hosting novel quantum phases such as Bose–Einstein condensate[11], spin superfluidity[12], and topological magnonic insulators[13].

Magnetic heterostructures can provide more functionalities and richer properties because exchange interactions between different layers provide another control knot [14,15]. While ferromagnetic metal-based heterostructures have been extensively studied and applied in commercial devices such as magneto-resistive random-access memory [15], magnetic insulator-based heterostructures are still on the horizon yet already showcased a few promises, including strong interfacial couplings [16–20], magnon valve effects[21–23], control of magnon transport in the magnetic insulator layer using another magnetic layer[24,25], magnonic crystal [26], coherent magnon-magnon coupling [27–30], and topological spin textures [10]. Magnetic insulator heterostructures are also theoretically predicted to host exotic quantum phase such as magnon flat band [31]. However, to date, coherent magnon-magnon coupling has only been studied in hybrid systems consisting of a low damping YIG and another ferromagnetic metal [27–30]. The demonstration of low damping and strong coherent coupling in purely magnetic insulator bilayers is lacking.

In this work, we demonstrate ultralow damping and strong magnon-magnon coupling in a TmIG/YIG heterostructure. We characterize the structural and magnetic properties of our TmIG/YIG heterostructures on gadolinium gallium garnet ($Gd_3Ga_5O_{12}$, GGG) using high-angle annular dark-field scanning transmission electron microscopy (HAADF-STEM), X-ray diffraction (XRD), and vibrating sample magnetometry (VSM). Then, we investigate the magnetic dynamics in these bilayers by using a broadband ferromagnetic resonance (FMR) technique. We observe a strong coupling between the Kittel mode of YIG and perpendicular standing spin wave (PSSW) mode of TmIG. By matching the experimental FMR spectra with analytical calculations and micromagnetic simulations, we obtain the exchange coupling strength at the interface, which is dependent on the magnetic insulator layer thickness and coupling mode. Finally, we benchmark the dissipation rates and cooperativity in our samples against these in ferromagnetic metal-based heterostructures.

We prepare our TmIG/YIG on GGG substrates using pulsed laser deposition (see Methods). Atomic images from HAADF-STEM show a single crystallinity and perfect interfaces at the YIG/GGG and TmIG/YIG boundaries (Fig. 1a). Elemental mapping (Fig. 1b) proves there is no interdiffusion between different layers. Fig. 1c presents the high-resolution XRD spectra of TmIG/GGG, YIG/GGG, and TmIG/YIG/GGG bilayer films measured with the scattering vector normal to the <001> oriented cubic substrate. Along the sharp <004> peaks from the GGG substrate, the XRD spectra shows Laue oscillations, indicating a smooth surface and interface. We also measured the magnetic hysteresis loops for YIG, TmIG, and TmIG/YIG samples to quantify their saturation magnetizations (see Supplementary Note 1). In principle, exchange coupling strength (J) between different layers can be estimated from major and minor hysteresis loops [15]. We can estimate the interfacial J at the CoFeB(50 nm)/ TmIG(350 nm) interface is $-0.031\ \mathrm{mJ/m^2}$, indicating an antiferromagnetic exchange coupling (see Supplementary Note 1). However, YIG and TmIG have very similar coercive fields, preventing us from obtaining the coupling strength directly from the hysteresis loop measurements.

We measure the magnetization dynamics in TmIG(200 nm)/YIG(200 nm) bilayers using a field modulated FMR technique (see Methods). We mount the sample on a coplanar waveguide and apply a microwave current that generates radiofrequency magnetic fields (Fig. 2a). The absorption coefficient exhibits a peak when the FMR conditions for YIG and TmIG are met (Fig. 2b). We experimentally extract the resonance frequency at a specific field by fitting the frequency scan at the field using Lorentz functions (Fig. 3a). In addition to regular FMR peaks, we also observe anti-crossing at specific field-frequency points, which are signatures of exchange interaction-driven coupling of Kittel mode in YIG and PSSW modes in TmIG. To identify the underlying magnon modes responsible for the coupling, we list the formula of generalized excited spin wave modes in two layers ($\frac{\omega_i}{2\pi}$ $or$ $f_i$):

$$\frac{\omega_i}{2\pi} = f_i = \frac{\gamma_i}{2\pi}\sqrt{\left(\mu_0 H_{\text{ext}} + \frac{2A_{\text{ex},i}}{M_{s,i}}k_i^2\right)\left(\mu_0 H_{\text{ext}} + \frac{2A_{\text{ex},i}}{M_{s,i}}k_i^2 + \mu_0 M_{s,i}\right)},\ (1)$$

where $i$=YIG or TmIG, $\frac{\gamma_i}{2\pi} = (g_{\text{eff,i}}/2) \times 28$ GHz/T is the gyromagnetic ratio, $\mu_0$ is the permeability, $H_{\text{ext}}$ is the external field, $M_s$ is the effective magnetization, $A_{\text{ex}}$ is the exchange stiffness, and $k$ is the wavevector of the excited spin wave. Note that if there is no exchange interaction between YIG and TmIG, $k = \frac{n\pi}{d}$, where n is an integer and $d$ is the thickness of the magnetic insulator. By fitting the Kittel mode with n=0, we get $g_{\text{eff,YIG}} = 2$ ($\mu_0 M_{s,YIG} = 0.25$ T) and $g_{\text{eff,TmIG}} = 1.56$ ($\mu_0 M_{s,TmIG} = 0.24$ T) for YIG and TmIG, respectively, which are consistent with the previous report [34]. Then, by assuming zero exchange interaction and matching $\omega_{\text{YIG}} = \omega_{\text{TmIG}}$ from Eq. (1), we can understand the first (second) anti-crossing shown in Fig. 2b is from the coupling between n=0 mode in YIG and n=1 (n=2) mode in TmIG. A schematic of n=0 mode in YIG and n=1 mode in TmIG is shown in Fig. 2a. In addition, we determine the exchange stiffness of the TmIG to be 2.69 pJ/m, which is consistent with the previous report [35]. When there is an exchange interaction between YIG and TmIG, we expect an anti-crossing gap, which can be described by the minimum frequency separation of 2g. However, with only Eq. (1) the relation between the exchange interaction and the g value cannot be uniquely determined.

To fully understand the exchange coupling-driven magnon-magnon coupling, we perform the comprehensive numerical analysis and micromagnetic simulations (see Methods). We consider the boundary conditions at the interface and two surfaces of the TmIG/YIG bilayers and arrive at the formula (see Supplementary Note 2):

$$\frac{2A_{\text{ex,YIG}}}{M_{\text{s,YIG}}} k_{\text{YIG}} \tan(k_{\text{YIG}} d_{\text{YIG}}) \cdot \frac{2A_{\text{ex,TmIG}}}{M_{\text{s,TmIG}}} k_{\text{TmIG}} \tan(k_{\text{TmIG}} d_{\text{TmIG}}) = \frac{2J}{\mu_0(M_{s,\text{YIG}} + M_{s,\text{TmIG}})}\left[\frac{2A_{\text{ex,YIG}}}{M_{\text{s,YIG}}} k_{\text{YIG}} \tan(k_{\text{YIG}} d_{\text{YIG}}) + \frac{2A_{\text{ex,TmIG}}}{M_{\text{s,TmIG}}} k_{\text{TmIG}} \tan(k_{\text{TmIG}} d_{\text{TmIG}})\right],\ (2)$$

where $J$ is the interfacial exchange coupling strength. By solving $\omega_{\text{YIG}} = \omega_{\text{TmIG}}$ from Eq. (1) and Eq. (2) together, we can get a set of ($k_{\text{YIG}}$, $k_{\text{TmIG}}$) values that correspond to different modes. In the presence of exchange interaction, $k$ will not be precisely equal to $\frac{n\pi}{d}$ anymore. As a result, the degeneracy is lifted at the crossing point and we have two frequencies corresponding to two ($k_{\text{YIG}}$, $k_{\text{TmIG}}$) values. By employing $J = -0.057$ mJ/m$^2$ (see Supplementary Table 1), we have obtained high consistency between the experimental and calculated spectra of field-frequency points in the entire range (Fig. 3a). The negative sign suggests an antiferromagnetic exchange coupling between TmIG and YIG. The strength is also comparable with the ferromagnetic metal/YIG bilayers [27]. We have also carried the FMR measurement on the reference TmIG(350 nm)/CoFeB(50 nm) sample (see Supplementary Note 4). We get $J = -0.032$ $m$J/m$^2$, which is close to the result from the VSM loop measurements. This consistency suggests that we can reliably extract $J$ values of TmIG/YIG samples from the FMR measurement.

We further study the thickness and mode dependence of anti-crossing gap (2g). We extract the g value from the frequency scan, for example, g = 85 MHz for the TmIG(200 nm)/YIG(200 nm) bilayer (Fig. 3b). We find the gap reduces as the layer thickness increases (Fig. 3c). To understand this, we derive the approximate solution (see Supplementary Note 3):

$$g \approx \frac{\gamma_{YIG}\gamma_{TmIG}}{4\pi^2}\frac{J}{(M_{s,YIG}+M_{s,TmIG})}\cdot\frac{\sqrt{(2\mu_0 H_{\text{res}}+\mu_0 M_{s,YIG})(2\mu_0 H_{\text{res}}+\mu_0 M_{s,TmIG})}}{f_{res}}\cdot\frac{1}{\sqrt{d_{YIG}d_{TmIG}}}, \quad (3)$$

where $\omega_{\text{res}}$ and $H_{\text{res}}$ are the resonance frequency and field in the gap center, respectively. The calculated results show the same trend as in the experiments (Fig. 3c). Also, Eq. (3) allows us to analyze the g value for coupling of the YIG Kittel mode to different TmIG PSSW modes. We compare the experimental and calculated g values for the coupling of n=0 mode in YIG and n=2 mode in TmIG in Fig. 3c, where we confirm that the higher mode coupling results in a lower g in our case.

Finally, to evaluate the coupling cooperativity in TmIG/YIG bilayers, we have determined the individual dissipation rates. We first get Gilbert damping factors for YIG and TmIG from field scans at different frequencies when they are not coupled (see Supplementary Note 4). The extracted damping factors are plotted in Fig. 3d, where we find a damping factor as low as 4.91 (±0.79) $\times 10^{-4}$ in the 350 nm-thick TmIG. We also extract the dissipation rates for YIG and TmIG from frequency scans at different fields when they are not coupled (see Supplementary Note 4). As an example, $\kappa_{YIG} = 10\ MHz$ and $\kappa_{TmIG} = 29.5\ MHz$ for the TmIG(200 nm)/YIG(200 nm) bilayer. Therefore, $g > \kappa_{YIG}$ , $\kappa_{TmIG}$ , and $C = \frac{g^2}{\kappa_{YIG}\kappa_{TmIG}} = 24.5$ , concluding a strong coupling in the bilayer. In Fig. 4, we summarize the dissipation rates and cooperativity for TmIG- and ferromagnetic metal-based heterostructures that show magnon-magnon coupling. The TmIG has a very low dissipation rate compared to ferromagnetic metals, which is consistent with the ultralow Gilbert damping.

In summary, we demonstrate ultralow damping and dissipation rates in the TmIG and achieve strong magnon-magnon coupling and high cooperativity in the TmIG/YIG bilayers. The combined experimental and theoretical analyses allow us to determine the interfacial exchange coupling strengths in our all-insulator bilayers. The all-magnetic-insulator bilayers allow us to achieve ultralow damping insulating synthetic antiferromagnets, magnonic crystals, and other artificial structures to realize energy-efficient spin wave devices. Besides, the strong coupling between two distinct magnetic insulators with ultralow damping allows to explore the novel quantum phases, such as topological magnon insulators and magnon flatband.

## Figures and Captions

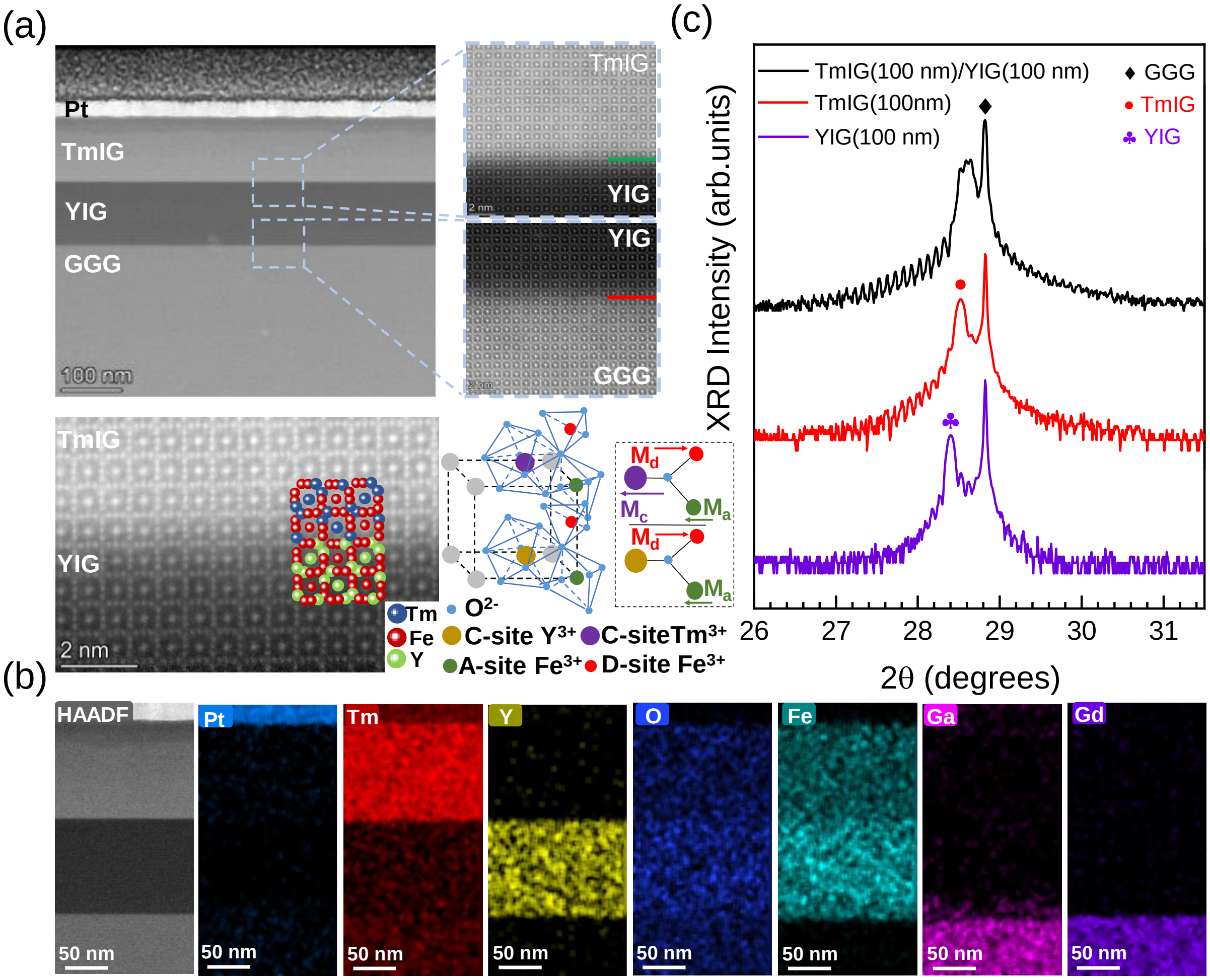

**Figure 1. Structural characterizations of YIG/TmIG heterostructures on GGG substrates. a,** High-angle annular dark-field scanning transmission electron microscopy (HAADF-STEM) image for the YIG/TmIG heterostructure on GGG substrates. The YIG/GGG and YIG/TmIG interfaces are denoted by the green and red lines in partial enlarged figure, respectively. Pt is used as a capping layer to prevent damage when preparing TEM samples. The inset of **a** demonstrates that two 1/8 of unit cells corresponding to YIG and TmIG at the interface, which have a garnet-type structures $(C)_3[A]_2(D)_3O_{12}$. **b,** Energy dispersive x-ray (EDX) spectra of different elements in the TmIG/YIG/GGG heterostructure. The images were taken along the <100> direction of the GGG substrate, and the distribution of elements is marked in the figure with color, respectively. **c,** High-resolution of X-ray diffraction spectra for the YIG(100 nm)/GGG, TmIG(100 nm)/GGG, and TmIG(100 nm)/YIG(100 nm)/GGG samples.

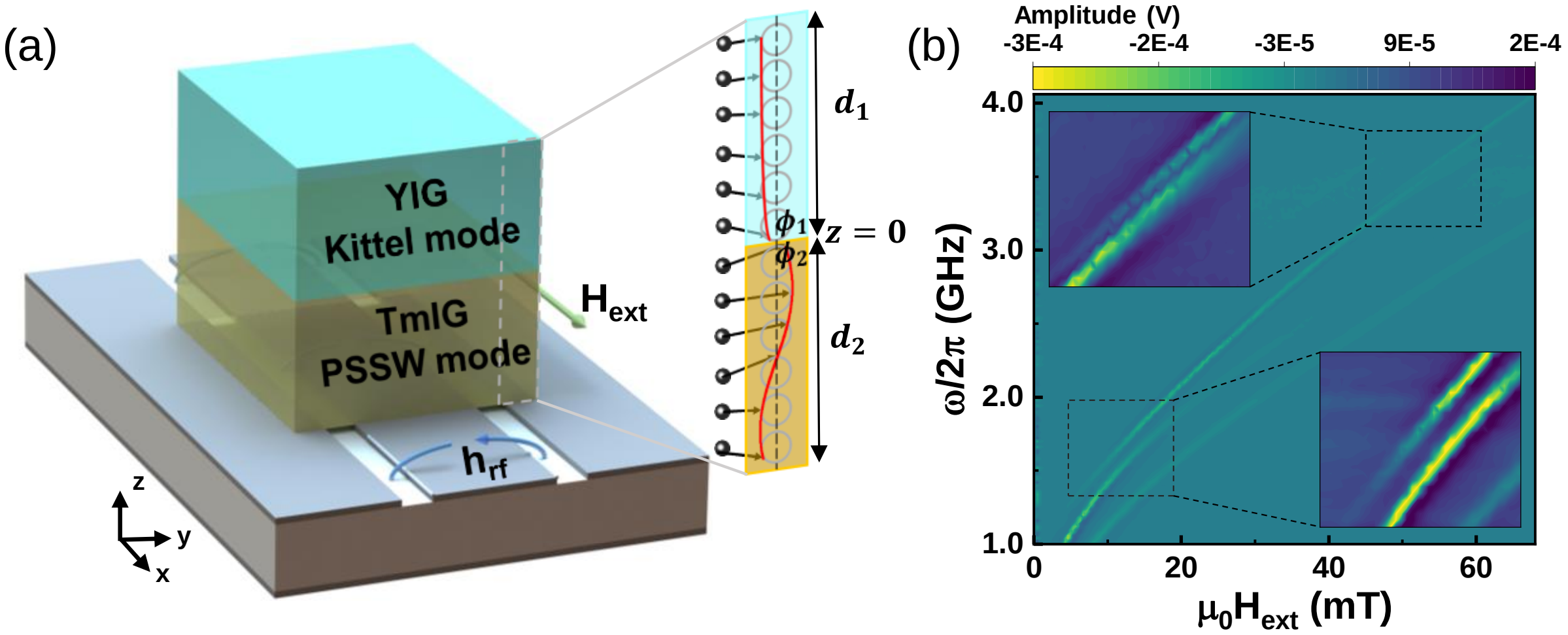

**Figure 2. Schematic diagram of the spin waves in the heterostructure and the measured resonance spectra. a,** Schematic illustration of the measurement set-up, where $h_{rf}$ and $H_{ext}$ stand for the microwave magnetic field and external static magnetic field, respectively. Spin-wave spectra are obtained by placing the sample face-down on a coplanar waveguide (CPW). The inset depicts the Kittel uniform spin wave mode in the YIG and the perpendicular standing spin wave (PSSW) mode in the TmIG. **b,** Experimentally color-coded spin-wave absorption spectra of the YIG(200 nm)/TmIG(200 nm) for the first three resonance modes of TmIG (n=0, 1, 2) and the uniform mode of YIG (n=0).

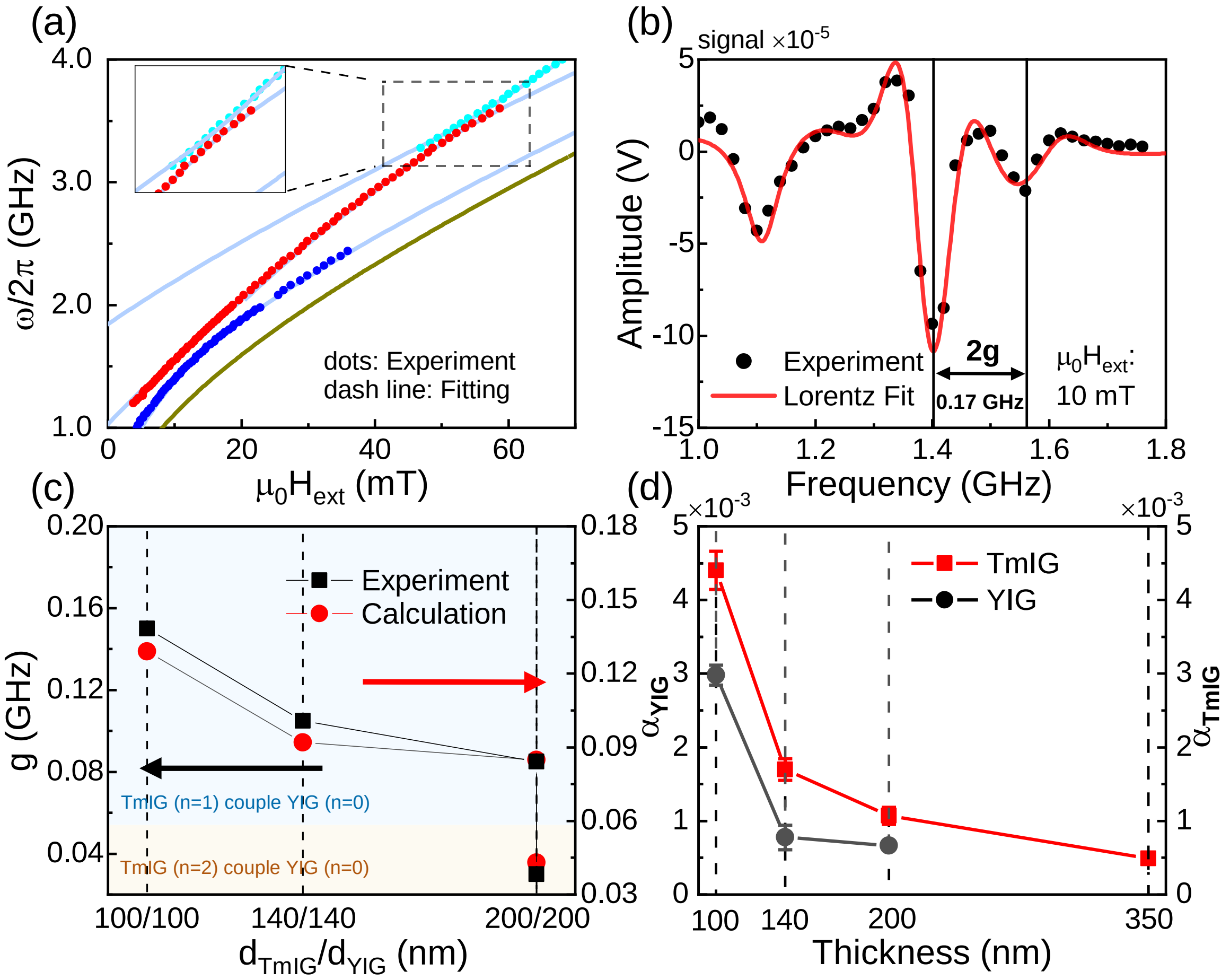

**Figure 3. Observation of strong magnon-magnon coupling and ultralow damping in YIG/TmIG bilayers. a,** Resonant absorption peaks of the two hybrid modes as a function of external magnetic field with the YIG(200 nm)/TmIG(200 nm) bilayer. Solid curves show the numerical theory method fitting as hybrid modes. Data points are extracted from experimental data by reading out the minimum of each resonant peak from Fig. 2(b). **b,** Spin wave spectra at minimum resonance separation ($\mu_0 H_{ext} = 10$ mT) in magnetic insulator bilayers with the YIG(200 nm)/TmIG(200 nm) bilayer. **c,** Coupling strength g between TmIG (n=1,2) mode and YIG (n=0) mode as a function of the YIG thickness. Red circles are experimental results and black squares are from theoretical calculations. Red dots: Experiments. **d,** Thickness dependence of Gilbert damping factors of YIG and TmIG in the YIG/TmIG bilayers.

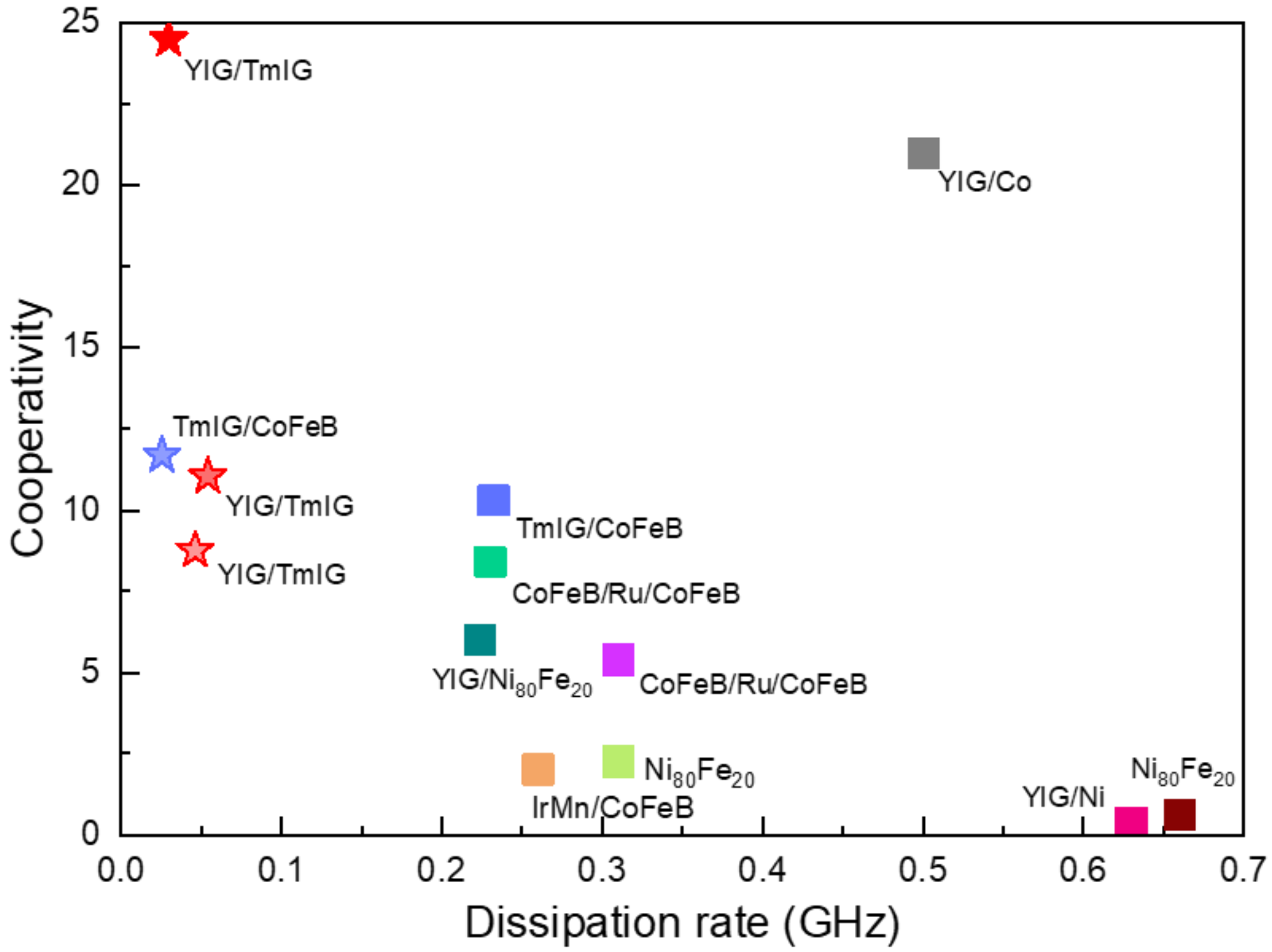

**Figure 4. Summary of dissipation rates in TmIG and ferromagnetic metals versus cooperativities in TmIG- and ferromagnetic metal-based heterostructures.** Star and square symbols are dissipation rates for TmIG and ferromagnetic metals, respectively. All TmIG-related results are from this work and other data points are from refs. [27,29,36–40] (see Supplementary Table 2 for details).

**References**

1. Chumak, A. V, Vasyuchka, V. I. I., Serga, A. A. A. & Hillebrands, B. Magnon spintronics. *Nat. Phys.* **11**, 453 (2015).

2. Khitun, A., Bao, M. & Wang, K. L. Magnonic logic circuits. *J. Phys. D. Appl. Phys.* **43**, 264005 (2010).

3. Pirro, P., Vasyuchka, V. I., Serga, A. A. & Hillebrands, B. Advances in coherent magnonics. *Nat. Rev. Mater.* **6**, 1114–1135 (2021).

4. Kruglyak, V. V, Demokritov, S. O. & Grundler, D. Magnonics. *J. Phys. D. Appl. Phys.* **43**, 264001 (2010).

5. Cornelissen, L. J., Liu, J., Duine, R. A., Youssef, J. Ben & Van Wees, B. J. Long-distance transport of magnon spin information in a magnetic insulator at room temperature. *Nat. Phys.* **11**, 1022 (2015).

6. Uchida, K. *et al.* Spin Seebeck insulator. *Nat. Mater.* **9**, 894–897 (2010).

7. Avci, C. O. *et al.* Current-induced switching in a magnetic insulator. *Nat. Mater.* **16**, 309–314 (2017).

8. Shao, Q. *et al.* Role of dimensional crossover on spin-orbit torque efficiency in magnetic insulator thin films. *Nat. Commun.* **9**, 3612 (2018).

9. Shao, Q. *et al.* Topological Hall effect at above room temperature in heterostructures composed of a magnetic insulator and a heavy metal. *Nat. Electron.* **2**, 182–186 (2019).

10. Vélez, S. *et al.* Current-driven dynamics and ratchet effect of skyrmion bubbles in a ferrimagnetic insulator. *Nat. Nanotechnol.* **17**, 834–841 (2022).

11. Demokritov, S. O. *et al.* Bose–Einstein condensation of quasi-equilibrium magnons at room temperature under pumping. *Nature* **443**, 430–433 (2006).

12. Takei, S. & Tserkovnyak, Y. Superfluid Spin Transport Through Easy-Plane Ferromagnetic Insulators. *Phys. Rev. Lett.* **112**, 227201 (2014).

13. Chisnell, R. *et al.* Topological Magnon Bands in a Kagome Lattice Ferromagnet. *Phys. Rev. Lett.* **115**, 147201 (2015).

14. Duine, R. A., Lee, K.-J., Parkin, S. S. P. & Stiles, M. D. Synthetic antiferromagnetic spintronics. *Nat. Phys.* **14**, 217–219 (2018).

15. Zabel, H. & Bader, S. D. *Magnetic Heterostructures*. (Springer Berlin Heidelberg, 2007). doi:10.1088/0031-9112/23/4/020.

16. Gomez-Perez, J. M. *et al.* Synthetic Antiferromagnetic Coupling Between Ultrathin Insulating Garnets. *Phys. Rev. Appl.* **10**, 044046 (2018).

17. Becker, S. *et al.* Magnetic Coupling in Y3Fe5 O12/Gd3Fe5 O12 Heterostructures. *Phys. Rev. Appl.* **16**, 014047 (2021).

18. Kumar, R., Sarangi, S. N., Samal, D. & Hossain, Z. Positive exchange bias and inverted hysteresis loop in Y3 Fe5 O12/ Gd3 Ga5 O12. *Phys. Rev. B* **103**, 064421 (2021).

19. Liang, J. M. *et al.* Observation of Interfacial Antiferromagnetic Coupling between Ferrimagnetic Garnet Thin Films. *IEEE Trans. Magn.* 1–1 (2021) doi:10.1109/TMAG.2021.3087822.

20. Quarterman, P. *et al.* Probing antiferromagnetic coupling in magnetic insulator/metal heterostructures. *Phys. Rev. Mater.* **6**, 094418 (2022).

21. Wu, H. *et al.* Magnon Valve Effect between Two Magnetic Insulators. *Phys. Rev. Lett.* **120**, 097205 (2018).

22. Cramer, J. *et al.* Magnon detection using a ferroic collinear multilayer spin valve. *Nat. Commun.* **9**, 1089 (2018).

23. Guo, C. Y. *et al.* Magnon valves based on YIG/NiO/YIG all-insulating magnon junctions. *Phys. Rev. B* **98**, 134426 (2018).

24. Han, J. *et al.* Nonreciprocal Transmission of Incoherent Magnons with Asymmetric Diffusion Length. *Nano Lett.* **21**, 7037–7043 (2021).

25. Fan, Y. *et al.* Manipulation of Coupling and Magnon Transport in Magnetic Metal-Insulator Hybrid Structures. *Phys. Rev. Appl.* **13**, 061002 (2020).

26. Qin, H. *et al.* Nanoscale magnonic Fabry-Pérot resonator for low-loss spin-wave manipulation. *Nat. Commun.* **12**, 2293 (2021).

27. Li, Y. *et al.* Coherent Spin Pumping in a Strongly Coupled Magnon-Magnon Hybrid System. *Phys. Rev. Lett.* **124**, 117202 (2020).

28. Klingler, S. *et al.* Spin-Torque Excitation of Perpendicular Standing Spin Waves in Coupled YIG/Co Heterostructures. *Phys. Rev. Lett.* **120**, 127201 (2018).

29. Chen, J. *et al.* Strong Interlayer Magnon-Magnon Coupling in Magnetic Metal-Insulator Hybrid Nanostructures. *Phys. Rev. Lett.* **120**, 217202 (2018).

30. Qin, H., Hämäläinen, S. J. & van Dijken, S. Exchange-torque-induced excitation of perpendicular standing spin waves in nanometer-thick YIG films. *Sci. Rep.* **8**, 5755 (2018).

31. Cheng, C. *et al.* Magnon flatband effect in antiferromagnetically coupled magnonic crystals. *Appl. Phys. Lett.* **122**, 082401 (2023).

32. He, C. *et al.* Spin-Torque Ferromagnetic Resonance in W/Co - Fe - B/ W/Co - Fe - B/Mg O Stacks. *Phys. Rev. Appl.* **10**, 034067 (2018).

33. Kalarickal, S. S. *et al.* Ferromagnetic resonance linewidth in metallic thin films: Comparison of measurement methods. *J. Appl. Phys.* **99**, 093909 (2006).

34. Crossley, S. *et al.* Ferromagnetic resonance of perpendicularly magnetized Tm3Fe5O12/Pt heterostructures. *Appl. Phys. Lett.* **115**, (2019).

35. Rosenberg, E. R. *et al.* Magnetic Properties and Growth-Induced Anisotropy in Yttrium Thulium Iron Garnet Thin Films. *Adv. Electron. Mater.* **7**, 2100452 (2021).

36. Adhikari, K., Sahoo, S., Mondal, A. K., Otani, Y. & Barman, A. Large nonlinear ferromagnetic resonance shift and strong magnon-magnon coupling in N i80 F e20 nanocross array. *Phys. Rev. B* **101**, 054406 (2020).

37. Adhikari, K., Choudhury, S., Barman, S., Otani, Y. & Barman, A. Observation of magnon–magnon coupling with high cooperativity in Ni 80 Fe 20 cross-shaped nanoring array. *Nanotechnology* **32**, 395706 (2021).

38. Shiota, Y., Taniguchi, T., Ishibashi, M., Moriyama, T. & Ono, T. Tunable Magnon-Magnon

Coupling Mediated by Dynamic Dipolar Interaction in Synthetic Antiferromagnets. *Phys. Rev. Lett.* **125**, 017203 (2020).

39. Wang, H. *et al.* Hybridized propagating spin waves in a CoFeB/IrMn bilayer. *Phys. Rev. B* **106**, 064410 (2022).

40. Hayashi, D. *et al.* Observation of mode splitting by magnon–magnon coupling in synthetic antiferromagnets. *Appl. Phys. Express* **16**, 053004 (2023).

## Methods

### Material growth

High-quality epitaxial YIG (lattice constant 12.376Å), TmIG (lattice constant 12.324 Å), and YIG/TmIG thin films were deposited on <100>-oriented GGG (lattice constant 12.383 Å) single-crystal substrates using pulsed laser deposition (PLD). Prior to deposition, the substrates were cleaned with acetone, alcohol, and deionized water, followed by annealing in air at 1000°C for 6 hours. The films were deposited at 710°C in an oxygen atmosphere of 100 mTorr, with a base pressure of better than $2 \times 10^{-6}\ mTorr$, using a 248 nm KrF excimer laser with a repetition rate of 10 Hz. In-situ post-annealing was carried out at the deposition temperature for 10 minutes in 10 Torr oxygen ambient, followed by natural cooling to room temperature.

### Material characterizations

The cross-sectional (TmIG/YIG/GGG) TEM lamella with a thickness of ~70 nm was fabricated by the focused ion beam (FIB) technique in a Thermofisher Helios G4 UX dual beam system. The atomic resolution high-angle annular dark-field scanning transmission electron microscopy (HAADF-STEM) images viewed along <100> orientation and the EDS data were obtained using FEI Titan Themis G2 TEM with a probe corrector at an acceleration voltage of 300 kV. The C2 aperture is 70 µm and the camera length is 115 mm, corresponding to a probe convergence semi-angle of 23.9 mrad. The image filtering and EDS mapping analysis were processed using the Velox software.

Film thickness was measured using Surface Profiler (Bruker DektakXT). The microstructure of the samples was analyzed using X-ray diffractometry (SmartLab, Rigaku Co.) with Cu Ka1 radiation (λ = 1.5406 Å). Magnetic properties were assessed with vibrating sample magnetometers (Physical Property Measurement System, Quantum Design, or Lakeshore).

### Resonance studies

The magnetization dynamics measurement was performed using the field-modulation FMR technique at room temperature. During the measurement, the GGG/YIG/TmIG sample was mounted in the flip-chip configuration (TmIG side facing down) on top of the signal-line of a coplanar waveguide for broadband microwave excitation. An external bias field, H, was applied in-plane perpendicular to the rf field of the CPW. We used a modulation frequency of $\Omega/2\pi = 81.57$ Hz (supplied by a lock-in amplifier and provided by a pair of modulation coil) and a modulation field of about 1.1 Oe. The microwave signal was delivered from a signal generator (0 - 5 dBm) to one port of the board. The field-modulated FMR signal was measured from the other port by the lock-in amplifier in the form of a dc voltage, V, by using a sensitive rf diode. We swept the bias field, H, and at each incremental frequency f, to construct the V[f, H] dispersion contour plots.

### Micromagnetic simulations

Our finite-element model implements coupled LLG equation with antiferromagnetic interfacial exchange interaction in magnetic insulator heterostructure in order to simulate the strong magnon-magnon coupling in frequency-domain. The technical details are provided in Supplementary Note 2.

## Acknowledgements

The authors appreciate insightful discussions with S. S. Kim, Y. Tserkovnyak, Z. Zhang, A. Comstock, D. Sun, Y. Li. The sample fabrication, structural characterization, and data analysis at HKUST were supported by National Key R&D Program of China (Grants No.2021YFA1401500). The magnetic dynamics measurement and analysis were supported by the U.S. National Science Foundation under Grant No. ECCS-2246254. The thin film deposition work in PolyU through the GRF grant 15302320. The authors also acknowledge support from RGC General Research Fund (Grant No. 16303322), State Key Laboratory of Advanced Displays and Optoelectronics Technologies (HKUST), and Guangdong-Hong Kong-Macao Joint Laboratory for Intelligent Micro-Nano Optoelectronic Technology (Grant No. 2020B1212030010).

## Data availability

The data that support the plots within this paper and other findings of this study are available at XXX. (Authors' note: the data will be uploaded after the acceptance of this manuscript.)

## Author contributions

W. Z. and Q. S. conceived the idea. J. L. did the VSM, XRD, partial FMR measurements, and data analysis with help from X. W., S. K. C., Z. R., R. L., Z. C., F. P. N., and S. K. Kim. Y. X. and W.Z. did FMR measurements with help from A. C. J. L. and D. C.W. L. grew the samples. C. L. and X.X. Z. did the TEM. J. L. and Q. S. and W. Z. wrote the manuscript with help from other co-authors.

## Competing interests

The authors declare that they have no competing financial interests.

# Supplementary Information

**Jiacheng Liu**, et al.

## Supplementary Note 1. Hysteresis loops by vibrating sample magnetometer (VSM) measurement for TmIG, YIG, and TmIG/YIG samples and TmIG, CoFeB, and TmIG/CoFeB samples

In principle, due to the existence of antiferromagnetic coupling at the heterostructure interface, we could also measure the major and minor hysteresis loops through VSM to estimate the interfacial coupling energy (J). We prepared 2 series of samples to do a comparison, [TmIG(100 nm), YIG(100 nm), TmIG(100 nm)/YIG (100 nm)) and TmIG (350 nm), CoFeB(50 nm), TmIG(350 nm)/CoFeB(50 nm)]. Due to the extremely similar coercive fields of YIG and TmIG, we cannot get the J directly from hysteresis loop data in Supplementary Fig. 1. Then, we made another series of samples (TmIG, CoFeB, and TmIG/CoFeB) for comparison. Measuring the hysteresis loops of TmIG(350 nm)/CoFeB(50 nm) and individual layers in Supplementary Fig. 2a, we could clearly observe that the process of magnetization flipping (black solid line) from TmIG(350 nm)/CoFeB(50 nm), which is caused by the different coercivity between TmIG (light blue solid line) and CoFeB (red solid line). We noticed that no sharp switching (like arrow points A to B to C) of the CoFeB layer is visible but a smooth increase (like arrow points A to C) of the measured magnetic moment until the bilayer magnetization is saturated. This could be explained by a direct interfacial exchange coupling between TmIG and CoFeB magnetizations 32[1-3]. Process (1-5) in Supplementary Figs. 2b-c show a possible magnetization flipping process in an exchange coupled heterostructure at an external magnetic field.

Subsequently, to quantify the interfacial exchange coupling field and energy, we measured the minor hysteresis loops of TmIG(350 nm)/CoFeB(50 nm) in Supplementary Fig. 3a. In an ideal state, (2) and (3) state is similar, but the difference is that the moment of (3) is about to flip the spin of the TmIG layer, and the moment of (2) has not yet reached the flip condition. The -minor loop measured from the experiment is a C-A-B-C loop (Supplementary Fig. 3b), indicating the existence of antiferromagnetic interfacial coupling. If there is no interfacial coupling, the loop will be like E-A-D-E, because it only depends on its coercivity of TmIG, that is, the forward and reverse minor loops should basically coincide (just like a single TmIG hysteresis loop). The shift from the minor hysteresis loops is precisely because of the interfacial exchange coupling that the two layers are in an antiferromagnetic relationship such that the extra $H_{ex}$ wants to make spins parallel, which is why the flipping (3) will be performed faster by $(H_{ex} - H_{ext})$. So, we can estimate the interfacial exchange coupling J value from the minor loops: $J = \mu_0 H_{ex} \times \frac{\Delta M_s}{V} \times t = -0.394 \times 10^{-3}(T) \cdot 1.975 \times 10^{5}(A/m) \cdot 400 \times 10^{-9}(m) = -0.03113\, mJ/m^2$, where $H_{ex}$ is exchange field induced by the interfacial coupling, $\Delta M_s$ is the value of magnetization reduction in the shadow region (C-B-D-E loop) for additional exchange filed, $V = (350nm + 50nm) \cdot 5mm \cdot 5mm$ is volume of the heterostructure, t is thickness of the sample. After the analysis of the interfacial exchange coupling by the minor loops, we also extract the interfacial coupling energy of the TmIG (350nm)/CoFeB (50nm) sample by ferromagnetic resonance (FMR) measurement, which is described in detail in Supplementary Note 3.

## Supplementary Note 2. Numerical analysis and micromagnetic simulation for the strong magnon-magnon coupling induced by the interface exchange interaction

### Numerical analysis

To solve the eigenfrequency of the anti-crossing curve under the strong magnon-magnon coupling (Fig. 3a), we conduct numerical analysis and micromagnetic simulations on the bilayer heterostructure following the method in references [4,5]. We assume that magnetic insulator layer 1 (MI1) describes index 1, and magnetic insulator layer 2 (MI2) describes index 2. The interface of the MI1 and MI2 is at z=0 (Supplementary Fig. 4a).

**Dispersion relation:** for the magnetic system, Landau-Lifshitz-Gilbert (LLG) equation [Eq. (S1)] describes the intrinsic precession of the spin in the materials:

$$\frac{\partial \overrightarrow{\boldsymbol{M}_i}}{\partial t} = -\mu_0 \gamma_i \overrightarrow{\boldsymbol{M}_i} \times \overrightarrow{\boldsymbol{H}_{eff,i}} + \frac{\alpha_i}{M_{s,i}} \left( \overrightarrow{\boldsymbol{M}_i} \times \frac{\partial \overrightarrow{\boldsymbol{M}_i}}{\partial t} \right), \tag{S1}$$

where whole magnetic insulator bilayers subjected to the static magnetic field $\overrightarrow{\boldsymbol{H}_{ext}}$ in the x direction and the dynamic magnetic field $\overrightarrow{\boldsymbol{h}_{rf}}$ generated by the coplanar waveguide (CPW) in the y direction. Since the dynamic magnetic field induced by the microwave is so small $|\overrightarrow{\boldsymbol{h}_{rf}}| \ll |\overrightarrow{\boldsymbol{H}_{ext}}|$ that precessing amplitude of spin moment can be divided into static and dynamic parts:

$$\overrightarrow{\boldsymbol{M}_i} = M_{s,i}\left(\overrightarrow{\boldsymbol{m}_{0,i}} + \delta\overrightarrow{\boldsymbol{m}_i}\right), \overrightarrow{\boldsymbol{m}_{0,i}} = \begin{bmatrix} 1 \\ 0 \\ 0 \end{bmatrix}, \delta\overrightarrow{\boldsymbol{m}_i} = \delta\overrightarrow{\boldsymbol{m}_{0,i}}e^{j\omega t} = \begin{bmatrix} \delta m_{x,i} \\ \delta m_{y,i} \\ \delta m_{z,i} \end{bmatrix}, \tag{S2}$$

where $\overrightarrow{\boldsymbol{m}_0}$ is normalization constant modulo 1, which is the static expression of spin moment. $\delta\overrightarrow{\boldsymbol{m}_i}$ is micro-perturbations induced by microwave fields, and its amplitude is much smaller than $|\overrightarrow{\boldsymbol{m}_0}|$. $\overrightarrow{\boldsymbol{H}_{eff}}$ includes an external magnetic statistic magnetic field $\overrightarrow{\boldsymbol{H}_{ext}}$ [Eq. (S3)] in x direction, dynamic magnetic field $\overrightarrow{\boldsymbol{h}_{rf}}$, $|\overrightarrow{\boldsymbol{h}_{rf}}| \ll |\overrightarrow{\boldsymbol{H}_{ext}}|$, $\overrightarrow{\boldsymbol{h}_{rf}} \perp \overrightarrow{\boldsymbol{H}_{ext}}$, and demagnetization field $\overrightarrow{\boldsymbol{H}_{de,i}}$ [Eq. (S4)]. In particular, the sample is considered as a thin film, so $N_x = N_y = 0, N_z = 1$.

$$\overrightarrow{\boldsymbol{H}_{ext}} = \begin{bmatrix} H_x \\ 0 \\ 0 \end{bmatrix}, \overrightarrow{\boldsymbol{m}_{0,i}} \parallel \overrightarrow{\boldsymbol{H}_{ext}}, \tag{S3}$$

$$\overrightarrow{\boldsymbol{H}_{de,i}} = -\mathcal{N} \cdot \overrightarrow{\boldsymbol{M}_i} = -M_{s,i}\begin{bmatrix} N_x & 0 & 0 \\ 0 & N_y & 0 \\ 0 & 0 & N_z \end{bmatrix}\left(\overrightarrow{\boldsymbol{m}_{0,i}} + \delta\overrightarrow{\boldsymbol{m}_i}\right) = -\mu_0 M_{s,i}\begin{bmatrix} 0 \\ 0 \\ \delta m_{z,i} \end{bmatrix}, \tag{S4}$$

Substitute $\overrightarrow{\boldsymbol{M}_i}$ and $\overrightarrow{\boldsymbol{H}_{eff,i}}$ into Eq. (S1):

$$M_{s,i}\frac{\partial\left(\overrightarrow{\boldsymbol{m}_{0,i}} + \delta\overrightarrow{\boldsymbol{m}_i}\right)}{dt} = -\mu_0\gamma_i M_{s,i}\left(\overrightarrow{\boldsymbol{m}_{0,i}} + \delta\overrightarrow{\boldsymbol{m}_i}\right) \times \left(\overrightarrow{\boldsymbol{H}_{ext}} + \overrightarrow{\boldsymbol{H}_{de}} + \overrightarrow{\boldsymbol{h}_{rf}}\right)$$

$$+\frac{\alpha_i}{M_{s,i}}M_{s,i}\left(\overrightarrow{\boldsymbol{m}_{0,i}} + \delta\overrightarrow{\boldsymbol{m}_i}\right) \times M_{s,i}\frac{\partial\left(\overrightarrow{\boldsymbol{m}_{0,i}} + \delta\overrightarrow{\boldsymbol{m}_i}\right)}{\partial t}, \tag{S5}$$

where $\frac{\partial\overrightarrow{\boldsymbol{m}_{0,i}}}{\partial t} = 0$, and $\frac{\partial\delta\overrightarrow{\boldsymbol{m}_i}}{dt} = j\omega\delta\overrightarrow{\boldsymbol{m}_{0,i}}e^{j\omega t} = j\omega\delta\overrightarrow{\boldsymbol{m}_i}$. In Eq. (S5), $\overrightarrow{\boldsymbol{m}_{0,i}} \times \overrightarrow{\boldsymbol{H}_{ext}} = 0$ is due to

the magnetization and the static component of the external magnetic field are parallel to each other, $\delta\overrightarrow{\boldsymbol{m}_i} \times \overrightarrow{\boldsymbol{h}_{rf}} = 0$ is due to the second-order epsilon, and $\delta\overrightarrow{\boldsymbol{m}_i} \times \frac{\partial\delta\overrightarrow{\boldsymbol{m}_i}}{\partial t} = \delta\overrightarrow{\boldsymbol{m}_i} \times j\omega\delta\overrightarrow{\boldsymbol{m}_i} = 0$. So, Eq. (S5) could be simplified to:

$$j\omega\delta\overrightarrow{\boldsymbol{m}_i} = -\mu_0\gamma\left(\overrightarrow{\boldsymbol{m}_{0,i}} \times \overrightarrow{\boldsymbol{H}_{de}} + \overrightarrow{\boldsymbol{m}_{0,i}} \times \overrightarrow{\boldsymbol{h}_{rf}} + \delta\overrightarrow{\boldsymbol{m}_i} \times \overrightarrow{\boldsymbol{H}_{ext}}\right) + j\omega\alpha\left(\overrightarrow{\boldsymbol{m}_{0,i}} \times \delta\overrightarrow{\boldsymbol{m}_i}\right), \tag{S6}$$

The matrix form of Eq. (S6) is:

$$j\omega\begin{bmatrix}\delta m_{x,i}\\ \delta m_{y,i}\\ \delta m_{z,i}\end{bmatrix} = -\mu_0\gamma_i\left[\begin{bmatrix}1\\0\\0\end{bmatrix}\times\begin{bmatrix}0\\0\\-M_{s,i}\delta m_{z,i}\end{bmatrix}+\begin{bmatrix}1\\0\\0\end{bmatrix}\times\begin{bmatrix}h_x\\h_y\\h_z\end{bmatrix}+\begin{bmatrix}\delta m_{x,i}\\ \delta m_{y,i}\\ \delta m_{z,i}\end{bmatrix}\times\begin{bmatrix}H_x\\0\\0\end{bmatrix}\right] + j\omega\alpha_i\left(\begin{bmatrix}1\\0\\0\end{bmatrix}\times\begin{bmatrix}\delta m_{x,i}\\ \delta m_{y,i}\\ \delta m_{z,i}\end{bmatrix}\right), \tag{S7}$$

$$j\frac{\omega}{\mu_0\gamma_i}\begin{bmatrix}\delta m_{x,i}\\ \delta m_{y,i}\\ \delta m_{z,i}\end{bmatrix} = -\left[\begin{bmatrix}0\\M_{s,i}\delta m_{z,i}\\0\end{bmatrix}+\begin{bmatrix}0\cdot h_x\\-h_z\\h_y\end{bmatrix}+\begin{bmatrix}0\\H_x\delta m_z\\-H_x\delta m_y\end{bmatrix}\right]+j\frac{\omega\alpha_i}{\mu_0\gamma_i}\begin{bmatrix}0\\-\delta m_z\\ \delta m_y\end{bmatrix}, \tag{S8}$$

Simplify Eq. (S8) into $\vec{\boldsymbol{h}} = \chi^{-1}\delta\vec{\boldsymbol{m}}$:

$$\begin{bmatrix}0\cdot h_x\\h_y\\h_z\end{bmatrix} = \begin{bmatrix}j\frac{\omega}{\mu_0\gamma} & 0 & 0\\ 0 & \left(H_x + j\frac{\omega\alpha}{\mu_0\gamma}\right) & -j\frac{\omega}{\mu_0\gamma}\\ 0 & j\frac{\omega}{\mu_0\gamma} & \left(M_s + H_x - j\frac{\omega\alpha}{\mu_0\gamma}\right)\end{bmatrix}\begin{bmatrix}\delta m_{x,i}\\ \delta m_{y,i}\\ \delta m_{z,i}\end{bmatrix}, \tag{S9}$$

When the FMR condition is met, the system will reach maximum resonance, that means the system, $\vec{h} = 0, \chi^{-1}\vec{m} = 0$ has a non-zero solution [Eq. (S10)]:

$$|\chi^{-1}| = 0, \tag{S10}$$

$$\left(M_{s,i} + H_x - j\frac{\omega\alpha_i}{\mu_0\gamma_i}\right)\left(H_x + j\frac{\omega\alpha_i}{\mu_0\gamma_i}\right) - \frac{\omega^2}{\mu_0^2\gamma_i^2} = 0, \tag{S11}$$

$$(1-\alpha_i^2)\omega^2 - j\mu_0\gamma M_{s,i}\alpha_i\omega - \mu_0^2\gamma_i^2 H_x\left(H_x + M_{s,i}\right) = 0, \tag{S12}$$

$$\omega_{complex} = j\frac{\mu_0\gamma_i M_{s,i}\alpha_i}{2(1-\alpha_i^2)} + \frac{1}{2(1-\alpha_i^2)}\sqrt{\mu_0^2\gamma_i^2 M_{s,i}^2\alpha_i^2 + 4(1-\alpha_i^2)[\mu_0^2\gamma_i^2 H_{x,i}(H_{x,i}+M_{s,i})]}, \tag{S13}$$

When $\alpha \ll 1$, the Eq. (S13) will be simplified to:

$$\omega_i = \gamma_i\sqrt{\mu_0 H_x(\mu_0 H_x + \mu_0 M_{s,i})}, \tag{S14}$$

Eq. (S14) is Kittel mode by solving the linearized LLG equation with wave vector $k = 0$. If consider the exchange field $\overrightarrow{\boldsymbol{H}_{ex}} = \frac{2A_{ex,i}}{M_{s,i}}k_i^2$, the Eq. (S14) will be revised to:

$$\omega_{complex} = j\frac{\mu_0\gamma_i M_{s,i}\alpha_i}{2(1-\alpha_i^2)} + \frac{1}{2(1-\alpha_i^2)}\sqrt{\mu_0^2\gamma_i^2 M_{s,i}^2\alpha_i^2 + 4(1-\alpha_i^2)\gamma_i^2\left[\left(\mu_0 H_x + \frac{2A_{ex,i}}{M_{s,i}}k_i^2\right)\left(H_{x,i} + \frac{2A_{ex,i}}{M_{s,i}}k_i^2 + M_{s,i}\right)\right]} \tag{S15}$$

When $\alpha \ll 1$, the Eq. (S15) will be simplified to:

$$\omega_i = \gamma_i\sqrt{\left(\mu_0 H_x + \frac{2A_{ex,i}}{M_{s,i}}k_i^2\right)\left(\mu_0 H_x + \frac{2A_{ex,i}}{M_{s,i}}k_i^2 + \mu_0 M_{s,i}\right)}, \tag{S16}$$

where $k_i$ is the wave vector of the perpendicular standing spin wave (PSSW), and the propagation of the spin wave is in the z direction.

$$\omega_1 - \omega_2 = 0, \tag{S17}$$

For the magnetic insulator bilayer, Eq. (S15) with $\alpha$ and Eq. (S16) without $\alpha$ describe the dispersion relation within each layer. Eq. (S17) means that the necessary condition for the bilayer spin waves to exist simultaneously in the same external condition (same $h_{rf}$ and $H_{ext}$) in the two layers.

**Boundary conditions — free boundary condition and interface exchange boundary condition:** Now, we can focus on the boundary conditions in the magnetic bilayers. For the magnetic insulator bilayer system, free boundary condition should be applied to the top ($z = d_1$) [Eq. (S18)] and the bottom ($z = -d_2$) [Eq. (S19)] of the bilayer system, which also mean no pining at the surfaces:

$$\left.\frac{\partial \delta m_{z,1}}{\partial z}\right|_{z=d_1} = 0, \tag{S18}$$

$$\left.\frac{\partial \delta m_{z,2}}{\partial z}\right|_{z=-d_2} = 0, \tag{S19}$$

Considering that the dynamic magnetization in x and y directions is uniform, there is a wave vector ($k_\perp$) along the thickness direction. The spatial distribution of the PSSW can be expressed as:

$$\delta m_{z,i} = \delta m_{0,i} \cos(k_i z + \phi_i)\,, \tag{S20}$$

where $\phi_i$ is the phase of the spin wave. Substitute Eq. (20) into Eq. (S18, S19), we can obtain:

$$-k_1 d_1 + n_1 \pi = \phi_1, \tag{S21}$$

$$k_2 d_2 + n_2 \pi = \phi_2, \tag{S22}$$

Interface exchange energy J describes the number and strength of exchange bonds between magnetic insulator layer 1 and layer 2. The effect of exchange coupling energy $J\ (mJ/m^2)$ at the interface (z = 0) will be shown by the interface boundary conditions generated by the combination of the Huffman boundary conditions [6-8]. The conservation of magnetic energy flow at the interface leads to:

$$\left.\frac{2A_{ex,1}}{M_{s,1}}\frac{\partial \delta m_{z,1}}{\partial z} + \frac{2J}{\left(M_{s,1}+M_{s,2}\right)}\left(\delta m_{z,2} - \delta m_{z,1}\right)\right|_{z=0} = 0, \tag{S23}$$

$$\left.-\frac{2A_{ex,2}}{M_{s,2}}\frac{\partial \delta m_{z,2}}{\partial z} + \frac{2J}{\left(M_{s,1}+M_{s,2}\right)}\left(\delta m_{z,1} - \delta m_{z,2}\right)\right|_{z=0} = 0, \tag{S24}$$

The matrix form of Eqs. (S23, S24) is as follows:

$$\begin{bmatrix}0\\0\end{bmatrix} = \begin{bmatrix} -\frac{2J}{(M_{s,1}+M_{s,2})}\cos\phi_1 - \frac{2A_{ex,1}}{M_{s,1}}k_1 sin\phi_1 & \frac{2J}{(M_{s,1}+M_{s,2})}cos\phi_2 \\ \frac{2J}{(M_{s,1}+M_{s,2})}cos\phi_1 & -\frac{2J}{(M_{s,1}+M_{s,2})}\cos\phi_2 + \frac{2A_{ex,2}}{M_{s,2}}k_2 sin\phi_2 \end{bmatrix}\begin{bmatrix}\delta m_{0,1}\\ \delta m_{0,2}\end{bmatrix} \tag{S25}$$

The resonant condition requires that the determinant of the coefficient matrix of Eq. S25 vanishes:

$$\left[\frac{2J}{(M_{s,1}+M_{s,2})}\cos\phi_1+\frac{2A_{ex,1}}{M_{s,1}}k_1 sin\phi_1\right]\left[\frac{2J}{(M_{s,1}+M_{s,2})}\cos\phi_2-\frac{2A_{ex,2}}{M_{s,2}}k_2 sin\phi_2\right]$$
$$=\left[\frac{2J}{(M_{s,1}+M_{s,2})}cos\phi_2\right]\left[\frac{2J}{(M_{s,1}+M_{s,2})}cos\phi_1\right], \tag{S26}$$

Eq. (S26) is the relationship under the interface-exchanged boundary conditions, combined with the free boundary conditions of Eqs. (S21, S22), we can obtain:

$$\left[\frac{2J}{(M_{s,1}+M_{s,2})}-\frac{2A_{ex,1}}{M_{s,1}}k_1\,tan(k_1d_1)\right]\left[\frac{2J}{(M_{s,1}+M_{s,2})}-\frac{2A_{ex,2}}{M_{s,2}}k_2\,tan(k_2d_2)\right]$$
$$=\left[\frac{2J}{(M_{s,1}+M_{s,2})}\right]\left[\frac{2J}{(M_{s,1}+M_{s,2})}\right], \tag{S27}$$

$$\frac{2A_{ex,1}}{M_{s,1}}\mathrm{k}_1\tan(k_1d_1)\cdot\frac{2A_{ex,2}}{M_{s,2}}\mathrm{k}_2\tan(k_2d_2)$$
$$=\frac{2J}{(M_{s,1}+M_{s,2})}\left[\frac{2A_{ex,1}}{M_{s,1}}\mathrm{k}_1\tan(k_1d_1)+\frac{2A_{ex,2}}{M_{s,2}}\mathrm{k}_2\tan(k_2d_2)\right], \tag{S28}$$

To get the eigenfrequency, we can solve the transcendental equations of the dispersion condition [Eq. (S17)] and boundary condition [Eq. (S28)] for each external magnetic field $H_{ext}$ by traversal.

Although $f_1(k_1,k_2,H_{ext})=0$ [Eq. (S17)] and $f_2(k_1,k_2,H_{ext})=0$ [Eq. (S28)] cannot be explicitly functionalized, we can solve them numerically by handling them as implicit functions. Supplementary Fig. 4b shows the framework of the numerical analysis using MATLAB software. Next, we will solve the eigenfrequency in two cases $(J=0\ mJ/m^2,\ J\neq 0\ mJ/m^2)$:

When $J=0\ mJ/m^2$, transcendental Eq. (S28) will degenerate to:

$$k_1k_2\tan(k_1d_1)\tan(k_2d_2)=0, \tag{S29}$$

$k_1=n\pi/d_1$ or $k_2=n\pi/d_2$ is the solution of Eq. (S29), which means spin wave in magnetic insulator layer 1 and layer 2 is quantized as $n\pi/d$ and not influenced with each other. Substitute Eq. (S29) into Eq. (S16) (assuming that $\alpha\ll 1$):

$$\omega_i = \gamma_i \sqrt{\left(\mu_0 H_x + \frac{2A_{ex,i}}{M_{s,i}}\left(\frac{n\pi}{d}\right)^2\right)\left(\mu_0 H_x + \frac{2A_{ex,i}}{M_{s,i}}\left(\frac{n\pi}{d}\right)^2 + \mu_0 M_{s,i}\right)}, \tag{S30}$$

The resonant peaks in Supplementary Fig. 5a shows the FMR and PSSW modes of TmIG and YIG FMR mode as a function of magnetic field when $J = 0\ mJ/m^2$. We also show the process of solving $k_1, k_2$ by graphical method in Supplementary Figs. 5b-c. Without interfacial exchange energy, the hybrid modes will be degenerate to the uniform mode and PSSW modes, which also means that spin-wave vector is quantized with $n\pi/d$.

When $J \neq 0 = -0.0573 mJ/m^2$, we can clearly see the anti-crossing curve in the Supplementary Fig. 6a. A set of degenerate solutions of the shaded region in the Supplementary Fig. 6a are composed of a set of intersection points of Figs. 6b-c. It also needs to be noted that the evanescent wave emerges in the YIG layer when the frequency of the PSSW in TmIG layer is lower than the uniform mode in YIG layer ($k_1$ is pure imaginary part).

**Micromagnetic simulations based on finite-element model**

In addition to proposing a relatively complete theory, we show the strong magnon-magnon coupling regime by solving the LLG equation in the frequency domain based on COMSOL's micromagnetic simulation module [9]. We apply the frequency-domain LLG equations of Eq. (S31, S32) to the two magnetic thin films respectively:

$$-i\omega\delta\boldsymbol{m_i} = -\gamma_i \boldsymbol{m_i} \times \left(\boldsymbol{H_{eff}} + \boldsymbol{H_{ex}}\right) - i\omega\alpha_i \boldsymbol{m_i} \times \delta\boldsymbol{m_i} \tag{S31}$$

$$-i\omega\delta\boldsymbol{m_j} = -\gamma_j \boldsymbol{m_j} \times \left(\boldsymbol{H_{eff}} + \boldsymbol{H_{ex}}\right) - i\omega\alpha_j \boldsymbol{m_j} \times \delta\boldsymbol{m_j} \tag{S32}$$

where all parameters are consistent with the theory calculations. $H_{ex}$ is the effective filed-like torque induced by the interfacial exchange energy, which is defined as:

$$H_{ex} = \frac{2J}{M_i + M_j}\delta(z)\boldsymbol{m_j}, \int_{-\infty}^{+\infty}\delta(z) = 1 \tag{S33}$$

where $2J/(M_i + M_j)$ means that the average value of the interfacial exchange energy for different magnetization materials on both sides of the interface. $\delta(z)$ is impulse function, which shows that the interaction only occurs at the interface of the TmIG/YIG heterostructure. The simulation parameters to be set are shown in the following Supplementary Table S1.

To demonstrate the coupling strength between YIG and TmIG, we perform full-frequency full-magnetic field simulations. We characterize the strength of resonance through the response of $\delta m_z$ to magnetic field and microwave field. Through the simulation results shown in the Supplementary Fig. 7a, we can clearly see that TmIG (n=1,2) is coupled with YIG (n=0) to form anti-crossing at 105 Oe and at 470 Oe. Supplementary Fig. 7b shows the spectral resonance curve of the spin wave at 105 Oe. Then, we can see the internal dynamics of spin waves through simulation. To perform the resonance direction of the spin wave on the interface of the bilayers, we show the $\delta m_z$ distribution of the normalized intensity in the z direction for the two hybrid modes A and B (marked) in Supplementary Fig. 8.

## **<u>Supplementary Note 3. Thickness dependence of interfacial exchange coupling energy and coupling strength for TmIG/YIG samples</u>**

### **The sign judgment of interfacial exchange coupling energy**

Judgment 1: we obtained the interfacial coupling energy $(J < 0\, mJ/m^2)$ by fitting the experimental data with the theory of Supplementary Note 2. The values of the J are summarized in the Supplementary Table S2.

Judgment 2: We plot the spin-wave absorption spectra for TmIG(100 nm)/YIG(100 nm) in comparison with YIG(100 nm)/GGG (grey dashed line) in Supplementary Fig. 9. The lower resonant magnetic field $(H_{res})$ or higher resonant frequency $(\omega_{res})$ means that for the TmIG/YIG sample, the interfacial exchange field applied on YIG is opposite to the external magnetic field, which supports the antiferromagnetic coupling nature concluded in the main text. In the meantime, Eq. (S38) shows that $\delta k_{1(YIG)}^2$ has the same sign as J, which means if $J < 0\, mJ/m^2$, $\delta k_{1,YIG}^2 < 0$. According to the Eq. (S16), we can know the under the same external magnetic field, the resonant frequency required by TmIG/YIG sample is smaller than that of YIG/GGG, which is consistent with our results (Supplementary Fig.9). We could also roughly estimate the coupling strength $(J \approx H_{shift} \times \frac{M_{s1}+M_{s2}}{2} \times d_1 = -0.0042\,T \times 133.3\,kA/m \times 100nm = -0.0598\,mJ/m^2)$ by observing the shift [Eq. (S38)] in the FMR curve between the TmIG/YIG/GGG heterostructure and YIG/GGG single layer.

**Thickness dependence of interfacial coupling energy and coupling strength for TmIG/YIG samples**

**Interfacial coupling energy:** Supplementary Figs. 10a-c show the colormap of the spin-wave absorption spectra for the PSSW modes of TmIG and the uniform mode of YIG measured for TmIG(100 nm)/YIG(100 nm), TmIG(140 nm)/YIG(140 nm), and TmIG(200 nm)/YIG(200 nm) heterostructures. We fit the experimental data (Supplementary Figs. 10e-f) through numerical analysis in Supplementary Note 2. The results are summarized in the Supplementary Table 2.

**Coupling strength g:** We could also extract the magnon-magnon coupling strength, which is defined as the half of the minimal peak to peak frequency spacing in the anti-crossing induced by the interface exchange energy[10,11]. When the coupling strength is 0 $(J = 0\,mJ/m^2)$, the hybrid modes will be decoupled. The resonant magnetic field $(H_{ext})$ and resonant frequency $(f)$ where the minimum resonant separation is located should intersect, which means that $k_1 = 0, k_2 = n\pi/d_2$. At the existence of the interfacial exchange coupling energy $(J \neq 0\,mJ/m^2)$, we can obtain the perturbative solution. we make a difference to Eq. (S16) to get an expression of the coupling strength g $(\frac{\delta f}{2})$:

$$2\mathrm{f}_{\mathrm{res}}\delta f = \left(\frac{\gamma}{2\pi}\right)^2 \cdot \left[2(2\mu_0 H_{res} + \mu_0 M_s)\frac{2A_{ex}}{M_s}k\delta k + (\frac{2A_{ex}}{M_s})^2 4k^3 dk\right], \tag{S34}$$

For Eq. (S28), we can let $\frac{2A_{ex,1}}{M_{s,1}}\mathrm{k}_1 \tan(k_1 d_1)$ be A and $\frac{2A_{ex,2}}{M_{s,2}}\mathrm{k}_2 \tan(k_2 d_2)$ be B. So, the Eq. (S28) can be simplified:

$$1 - \frac{2J}{\left(M_{s,1} + M_{s,2}\right)}\left(\frac{1}{A} + \frac{1}{B}\right) = 0, \tag{S35}$$

We now consider $k_1 = 0 + \delta k_1, k_2 = n\pi/d_2 + \delta k_2$, $|\delta k_1| \ll 1\,, |\delta k_2| \ll 1$ as the perturbation solution corresponding to the minimum resonance separation of YIG's FMR mode and TmIG PSSW mode (n). Eq. (S35) will yield layer 1-dominated and layer 2-dominated resonance. For layer 1-dominated resonance, we have:

$$\frac{2J}{(M_{s,1} + M_{s,2})}\frac{1}{A} \approx 1 \text{ and } \frac{2J}{(M_{s,1} + M_{s,2})}\frac{1}{B} \ll 1, \quad (S36)$$

$$\frac{2J}{(M_{s,1} + M_{s,2})} \approx \frac{2A_{ex,1}}{M_{s,1}} k_1\, tan(k_1 d_1) = \frac{2A_{ex,1}}{M_{s,1}} k_1 \delta k_1 d_1 = \frac{2A_{ex,1}}{M_{s,1}} \delta k_1^2 d_1, \quad (S37)$$

$$\delta k_1^2 = \frac{2J}{(M_{s,1} + M_{s,2})}\frac{M_{s,1}}{2A_{ex,1}}\frac{1}{d_1}, \quad (S38)$$

For $k_1 = \delta k_1$, $(2\mu_0 H + \mu_0 M_s)\frac{2A_{ex}}{M_s} 2\delta \mathrm{k}_1 \gg \left(\frac{2A_{ex}}{M_s}\right)^2 4\delta k_1^3$

$$\delta f_1 \approx \left(\frac{\gamma_1}{2\pi}\right)^2 \left[2(2\mu_0 H_{res} + \mu_0 M_{s,1})\frac{2J}{(M_{s,1} + M_{s,2})}\frac{1}{d_1}\right]\frac{1}{2f_{res}}, \quad (S39)$$

For layer 2-dominated resonance, we have:

$$\frac{2J}{(M_{s,1} + M_{s,2})}\frac{1}{B} \approx 1 \text{ and } \frac{2J}{(M_{s,1} + M_{s,2})}\frac{1}{A} \ll 1, \quad (S40)$$

$$\delta k_2 = \frac{2J}{(M_{s,1} + M_{s,2})}\frac{M_{s,2}}{2A_{ex,2}}\frac{1}{n\pi/d_2}\frac{1}{d_2}, \quad (S41)$$

For $k_2 = \frac{n\pi}{d_2} + \delta k_2$, $(2\mu_0 H + \mu_0 M_s) \sim 10^{-1}, \frac{2A_{ex}}{M_s} \sim 10^{-18}, k_2 \sim \frac{n\pi}{d_2} \sim 10^7$, so we can get that $(2\mu_0 H + \mu_0 M_s)\frac{2A_{ex}}{M_s} 2k_2 \sim 10^{-12} \gg \left(\frac{2A_{ex}}{M_s}\right)^2 4k^3 \sim 10^{-15}$.

$$\delta f_2 \approx \left(\frac{\gamma_2}{2\pi}\right)^2 \left[2(2\mu_0 H_{res} + \mu_0 M_{s,2})\frac{2J}{(M_{s,1} + M_{s,2})}\frac{1}{d_2}\right]\frac{1}{2f_{res}}, \quad (S42)$$

$$\delta f^2 = \left(\frac{\gamma_1}{2\pi}\right)^2 \left(\frac{\gamma_2}{2\pi}\right)^2 \left(\frac{2J}{(M_{s,1} + M_{s,2})}\right)^2 \frac{[(2\mu_0 H_{res} + \mu_0 M_{s,1})(2\mu_0 H_{res} + \mu_0 M_{s,2})]}{f_{res}^2}\frac{1}{d_1 d_2}, \quad (S43)$$

$$g = \frac{\delta f}{2} = \frac{\gamma_1}{2\pi}\frac{\gamma_2}{2\pi}\frac{J}{(M_{s,1} + M_{s,2})}\frac{\sqrt{(2\mu_0 H_{res} + \mu_0 M_{s,1})(2\mu_0 H_{res} + \mu_0 M_{s,2})}}{f_{res}}\frac{1}{\sqrt{d_1 d_2}}, \quad (S44)$$

Where $g$ is the coupling strength, $H_{res}, f_{res}$ are the resonant magnetic field and frequency at the minimum resonance separation, $d_1, d_2$ is the thickness of the YIG layer (FMR mode) and TmIG layer (PSSW mode). The g value extracted in the experiment (Supplementary Figs. 11a-c) and calculation from Eq. (S44) are shown as Fig. 3c.

## Supplementary Note 4. The extraction of Gilbert damping, dissipation rate, and cooperativity from the FMR measurements

**Gilbert damping $\alpha$:** High coupling strength $g$ and extremely low Gilbert damping $\alpha$ are key factors to realize long-distance information transmission that is free of Joule heating. Gilbert damping refers to an intrinsic feature of magnetic substances which dictates the speed at which angular momentum is transferred to the crystal lattice, which is the key parameter that determines the spin-wave relaxation [13,14]. To demonstrate the superiority of low damping MI/MI bilayer system of this work, we focus on the line widths variation with the frequency of YIG/ TmIG heterostructure samples. The linewidth versus frequency experiment data with error bar for different thicknesses were extracted by fitting Lorentz equations in Supplementary Fig. 12. In strong coupling region, we can clearly observe that the line width value of the pink circle is significantly higher than that of the light blue circle, which is suggested that a coherent damping-like torque which acts along or against the intrinsic damping torque depending on the phase difference of the coupled dynamics of YIG and TmIG [4]; Away from the strong coupling region, we use $\Delta H = \Delta H_0 + \frac{4\pi\alpha}{\gamma} f$ for linear characterization, and the result is indicated by the dashed line in the Supplementary Fig. 12. The fitting effective damping constants are summarized in the Supplementary Table 2.

**Dissipation rate:** We extract the dissipation rate (half width at half maximum) from frequency scans at different fields when they are away from the coupling region (Supplementary Fig. 13). We fit the FMR mode of YIG and TmIG in Supplementary Fig. 13 through the Lorentz peak to get the dissipation rate.

**Cooperativity $C$:** The strong coupling implies coherent dynamics between the magnon and the magnon. In our case, all samples' $g > \kappa_1, g > \kappa_2$ mean that strong coupling is formed [11,12,15]. Through Lorentz peak fitting, we found that the dissipation rate will decrease as the thickness increases, which is consistent with the damping trend we calculated. Due to the magnetic insulator heterostructure, the dissipation rates in TmIG/YIG bilayers are particularly low compared with ferromagnetic metal-based heterostructures so that we can get a largest cooperativity $C = (g/\kappa_1)(g/\kappa_2) = 24.5$ in the TmIG(200 nm)/YIG(200 nm) case. The results are summarized in the Supplementary Table 2. As the thickness increases, the cooperativity increases significantly, mainly because the decrease rate of the dissipation rates is greater than that of coupling strength. The cooperativity is summarized in the Supplementary Table S2.

**Antiferromagnetic interfacial exchange coupling energy for the TmIG(350 nm)/CoFeB(50 nm) sample**

We also verify antiferromagnetic interfacial exchange coupling energy of the MI/FM TmIG(350 nm)/CoFeB(50 nm) by FMR modulation. We use theory (Supplementary Note 2) to analyze and fit the color-coded experiment data (Supplementary Figs. 14a-b) and get the TmIG/CoFeB interfacial coupling energy $J = -0.0321\, mJ/m^2$, showing that the FMR method is consistent with the minor hysteresis loop method ($J = -0.0311\, mJ/m^2$).

| Parameters | Symbol | Value |
|---|---|---|
| Mesh | $N$ | $20 \times 20 \times 50$ |
| Microwave field | $h_{rf}$ | $0.001T$ |
| Interfacial exchange energy | $J$ | $-0.0537\ mJ/m^2$ |
| YIG/TmIG gyromagnetic ratio | $\gamma_{YIG}/\gamma_{TmIG}$ | $28/21.81\ GHz/T$ |
| YIG/TmIG stiffness constant | $A_{ex,YIG}/A_{ex,TmIG}$ | $3.08/2.67\ pJ/m$ |
| YIG/TmIG Gilbert damping | $\alpha_{YIG}/\alpha_{TmIG}$ | $6.66 \times 10^{-4}/1.17 \times 10^{-3}$ |
| Frequency range | $(f_{start}, \Delta f, f_{end})$ | $(1, 0.01, 4)\ GHz$ |
| Magnetic field range | $(H_{start}, \Delta H, H_{end})$ | $(30, 5, 700)\ Oe$ |

**Supplementary Table S1**. **|** Parameters used to do the micromagnetic simulation with $TmIG_{200nm}$/ $YIG_{200nm}$ bilayer in Supplementary Figs. 7-8.

| Materials | $\kappa_{YIG}$ ($\alpha_{YIG}$) | $\kappa_{TmIG}$ ($\alpha_{TmIG}$) or $\kappa_{FM}$ ($\alpha_{FM}$) | g | C | J (mJ/m$^2$) |
|---|---|---|---|---|---|
| YIG(100 nm)/TmIG (100 nm) (this work) | 0.058 GHz (2.98e-3 ± 1.4e-4) | 0.046 GHz (4.4e-3 ± 3e-4) | 0.155 GHz | 8.76 | -0.0618 |
| YIG(140 nm)/TmIG (140 nm) (this work) | 0.0185 GHz (7.78e-4 ± 1.7e-4) | 0.054 GHz (1.69e-3 ± 2e-4) | 0.105 GHz | 11.04 | -0.0720 |
| YIG(200 nm)/TmIG (200 nm) (this work) | 0.01 GHz (6.64e-4 ± 4.2e-5) | 0.0295 GHz (1.07e-3 ± 1e-4) | 0.085 GHz | 24.49 | -0.0573 |
| TmIG(350 nm)/CoFeB (50 nm) (this work) | -- | $\kappa_{TmIG}$: 0.0255 GHz (4.91e-4±8e-5) or $\kappa_{FM}$: 0.232 GHz (3e-3±3.9e-5) | 0.263GHz | 11.69 | -0.0321 |
| YIG(20 nm)/Ni(20 nm)[29] * | 0.06GHz | 0.63 GHz | 0.12 GHz | 0.38 | -- |
| YIG(20 nm)/Co(30 nm)[29] * | 0.06GHz | 0.5 GHz | 0.79 GHz | 21 | -- |
| YIG(100 nm)/$Ni_{80}Fe_{20}$ (9 nm)[27] | 0.106GHz (2.3e-4) | 0.192 GHz (1.75e-3) | 0.35 GHz | 6 | -0.060 ± 0.011 |
| YIG(1 μm)/Co(50 nm) [28] | (7.2e-4 ± 3e-5) | (7.7e-3 ± 1e-4) | -- | -- | -- |
| $Ni_{80}Fe_{20}$(20 nm)[34] * | -- | 0.66 GHz | -- | 0.6 | -- |
| $Ni_{80}Fe_{20}$(20 nm)[35] * | -- | 0.31 GHz | -- | 2.25 | -- |
| IrMn(10 nm)/CoFeB (80 nm)[36] | -- | 0.26 GHz | -- | 2 | -- |
| CoFeB(15 nm)/Ru(0.6 nm) /CoFeB(15 nm)[37] * | -- | 0.31 GHz | -- | 5.39 | -- |
| CoFeB(15 nm)/Ru(0.6 nm) /CoFeB(15 nm)[38] * | -- | 0.23 GHz | -- | 8.4±1.3 | -- |

**Supplementary Table 2.** Summary of dissipation rates, damping factors, coupling strengths, cooperativities, and exchange interaction strengths in the studied bilayers and some reported YIG/ferromagnetic metal (FM) bilayers that show magnon-magnon coupling. All reference numbers are consistent with the main text. * In-plane confinement introduces extra dynamical dipolar coupling.

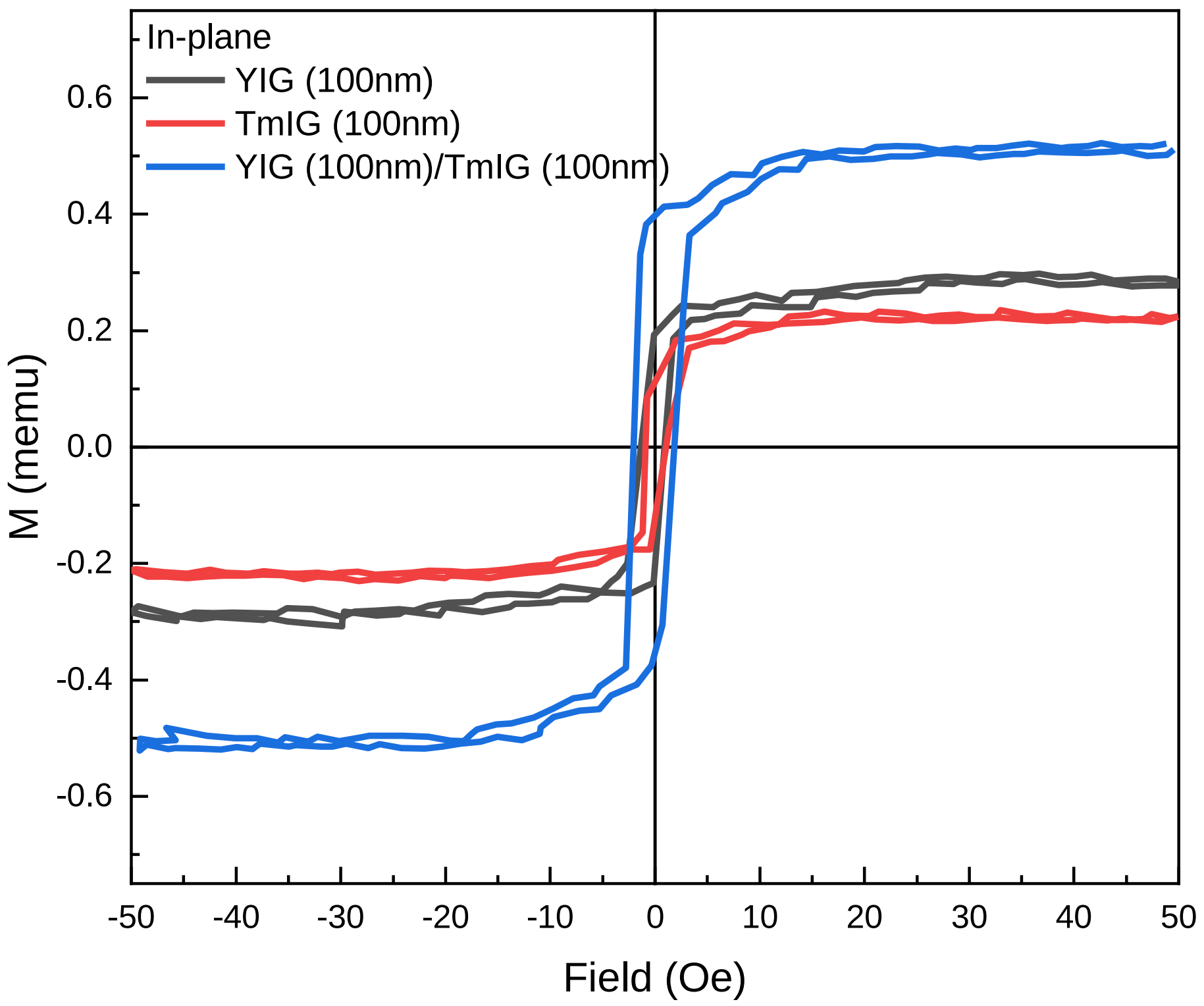

**Supplementary Figure 1 | Hysteresis loops for YIG(100 nm), TmIG(100 nm) and YIG (100nm)/TmIG(100nm) samples.** Since YIG and TmIG have approximately equal coercive fields (~1.5Oe), we could not obtain the interfacial coupling energy through the dynamic process of the switching of the two layers in the hysteresis loop.

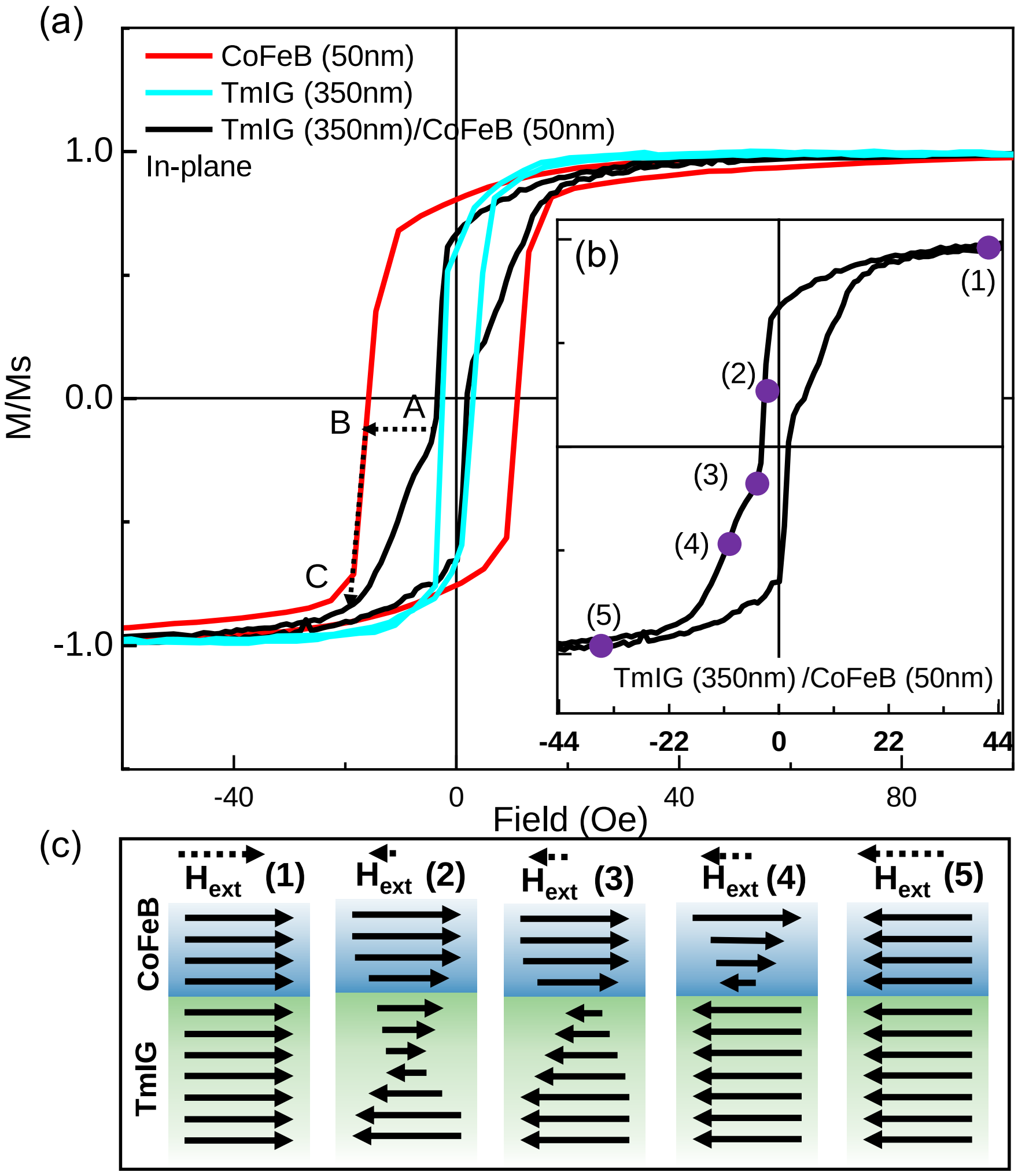

**Supplementary Figure 2 | Hysteresis loops for TmIG(350 nm), CoFeB(50 nm), and TmIG (350 nm)/CoFeB(50 nm) samples. a,** The individual magnetic thin film layers TmIG (light blue solid line) and CoFeB (red solid line) exhibit different coercivity, which could be directly observed the process of magnetization flipping between the two coercivity (black solid line). Furthermore, due to the existence of interfacial exchange coupling (J), no sharp switching (like the arrow point A to B to C) of the CoFeB layer is visible but a smooth increase (the arrow point A to C) of the measured magnetic moment until the bilayer magnetization is saturated. **b,** Enlarged detail of the TmIG(350 nm)/CoFeB(50 nm) heterostructure hysteresis loop. **c,** Process (1-5) shows a possible magnetization flipping in an exchange coupled heterostructure at an external magnetic field in **b**.

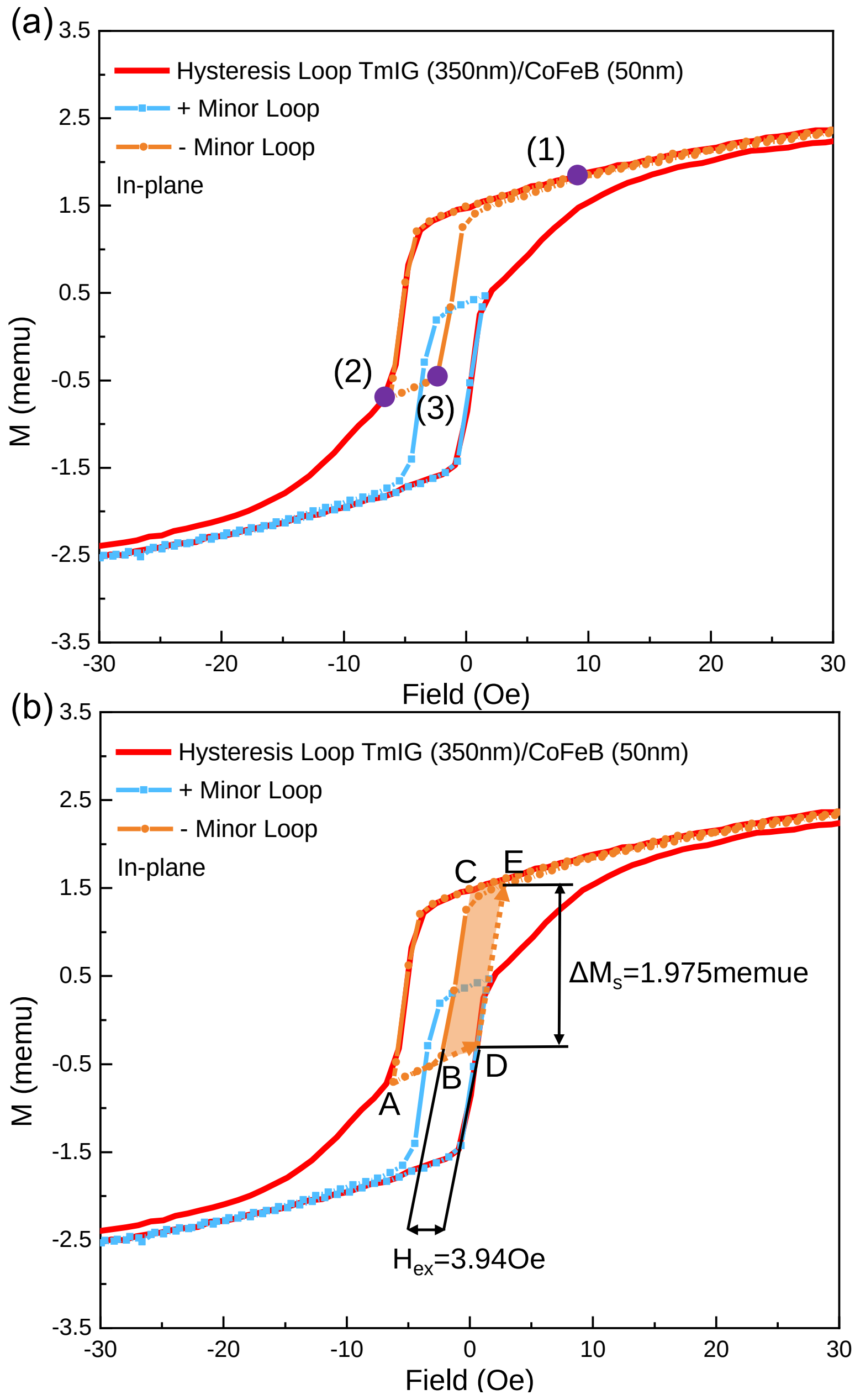

**Supplementary Figure 3 | Hysteresis loop and minor loops for TmIG(350 nm)/CoFeB (50 nm) sample. a,** Minor loops for TmIG(350 nm)/CoFeB(50 nm) sample. The non-coincidence of +minor loop and -minor loop indicates the existence of interface coupling. **b,** The process of analyzing interfacial exchange coupling using minor loops. The result of the experiment measurement is a C-A-B-C loop, indicating the existence of antiferromagnetic interfacial coupling. However, if there is no interfacial exchange coupling, the loop will be like E-A-D-E, because it only depends on its coercivity of TmIG, that is, the forward and reverse minor loops should basically coincide (just like individual TmIG Hysteresis loop). We can also get the

similar results from the + minor loop with the similar method.

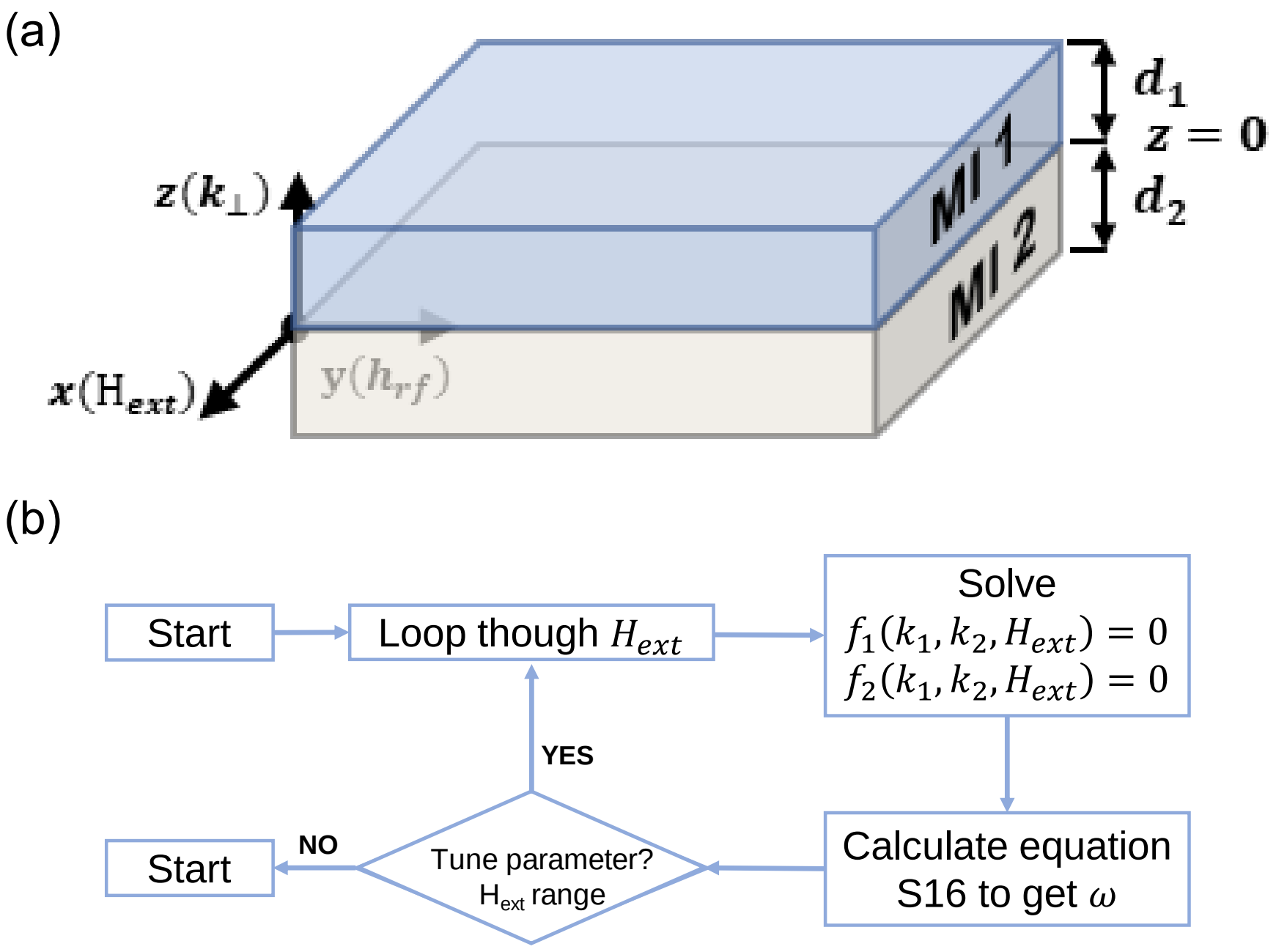

**Supplementary Figure 4 | Schematic of the magnetic insulator heterostructure with thickness of $d_1$ and $d_2$, respectively. a,** For the convenience of calculation, we set the interface of bilayer as the interface of z=0. The direction of the static external magnetic field ($H_{ext}$) is set as the x-axis, and the dynamic magnetic field ($h_{rf}$) generated by CPW is perpendicular to the static magnetic field on the y-axis, where $H_{ext} \gg h_{rf}$. **b,** Theoretical calculation framework from MATLAB. We traverse the value of $H_{ext}$ through the loop cycle and solve Eqs. (S17, S28) under each external magnetic field value to obtain the $k_1$ and $k_2$, then substitute them into the dispersion relationship to obtain the eigenfrequency.

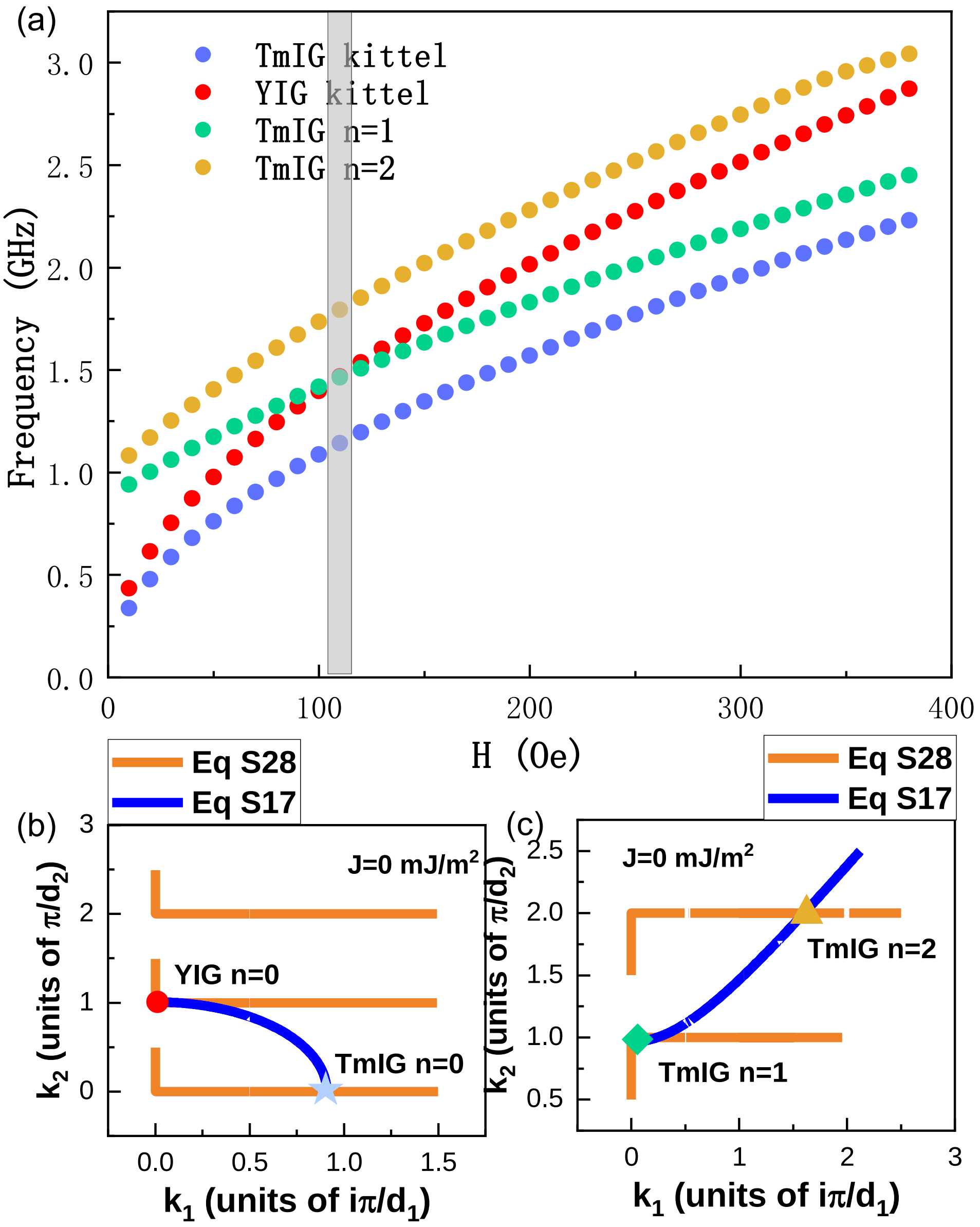

**Supplementary Figure 5 | Numerical solution for the eigenfrequency when $J = 0\ mJ/m^2$.** **a,** Frequencies of FMR and PSSW modes of TmIG(200 nm) and FMR mode of YIG(200 nm) as a function of H when $J = 0\ mJ/m^2$. **b-c,** Numerical solutions $k_1, k_2$ obtained by solving dispersion condition and boundary condition without interfacial exchange energy. When $J = 0\ mJ/m^2$, we could obviously see that the crossing points are degenerate solution in **a** from the shaded region, which indicates that $k_1, k_2$ is quantized solution with $n\pi/d_1$ and $n\pi/d_2$ respectively.

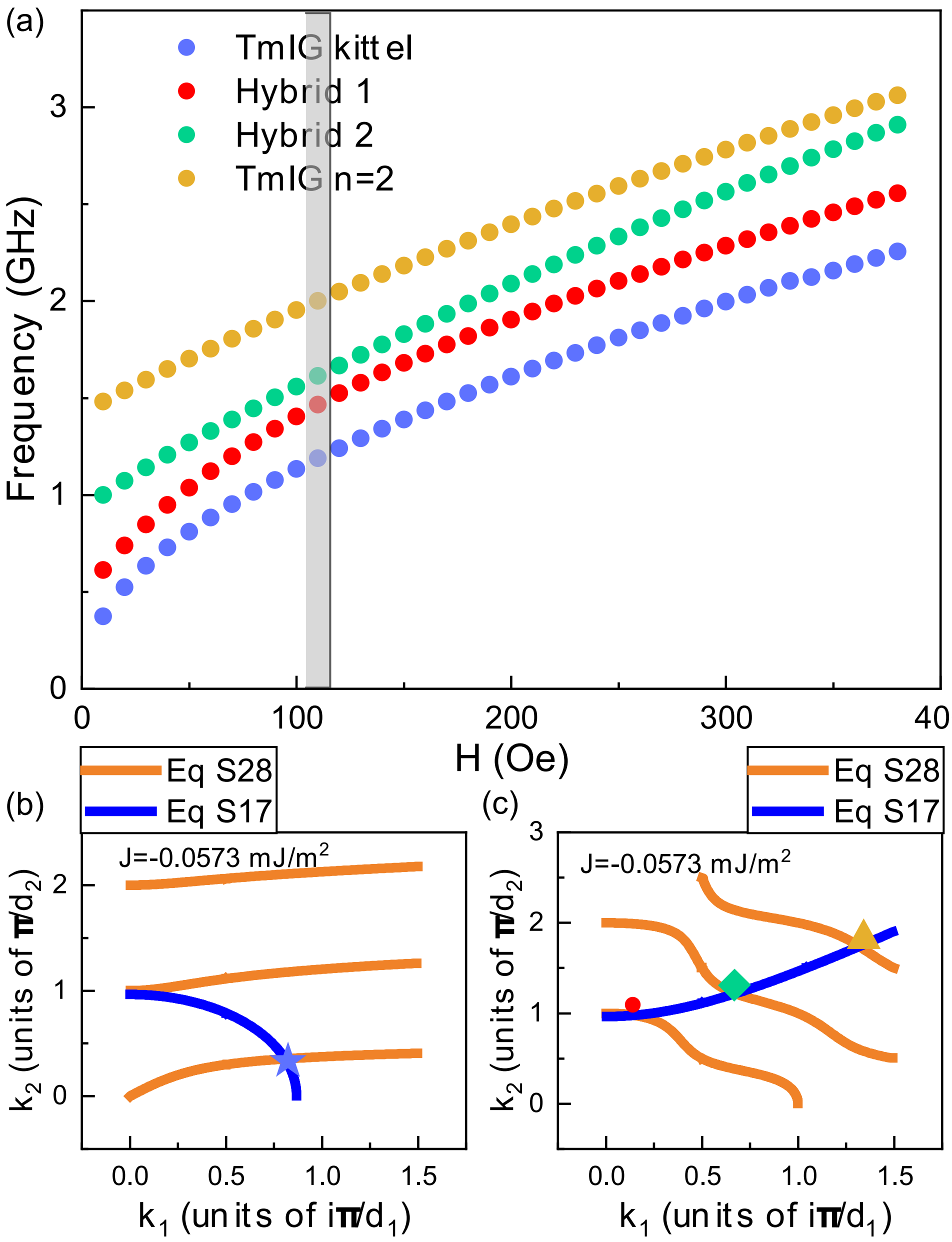

**Supplementary Figure 6 | Numerical solution for the eigenfrequency when $\boldsymbol{J = -0.0573 mJ/m^2}$.** **a,** Anti-crossing coupling frequencies of hybrid modes of TmIG(200 nm)/YIG(200 nm) as a function of H when $J = -0.0573\ mJ/m^2$. **b-c,** Numerical solutions $k_1, k_2$ obtained by solving dispersion condition and boundary condition without interfacial exchange energy. With the interfacial exchange energy, we could see that the spin-wave vector solution is no longer the integer of $n\pi/d$. It is noted that the evanescent wave emerges in the YIG layer when the frequency of the PSSW in TmIG layer is lower than the uniform mode in YIG layer ($k_1$ is pure imaginary part).

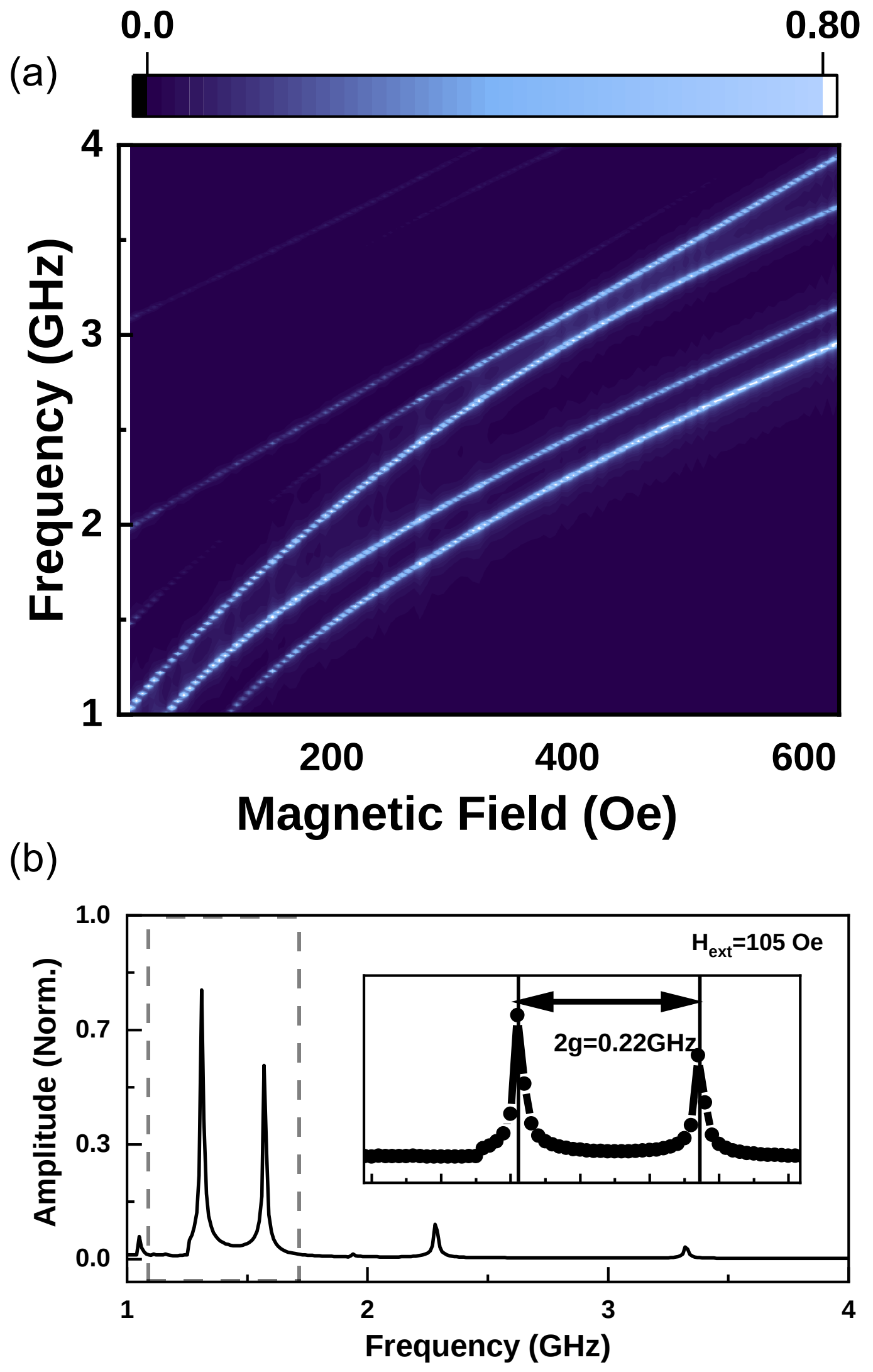

**Supplementary Figure 7 | Full micromagnetic simulation based on frequency domain. a,** Color-coded spin-wave spectra with frequency and magnetic field with the TmIG(200 nm)/ YIG(200 nm) heterostructure. we can clearly see that TmIG(n=1,2) is coupled with YIG(n=0) to form anti-crossing at 105 Oe and at 470 Oe. Then, we use the value of change in $\delta m_z$ as the peak response. **b,** The spin spectrum curve at the minimum resonance separation of the first anti-crossing (H=105 Oe).

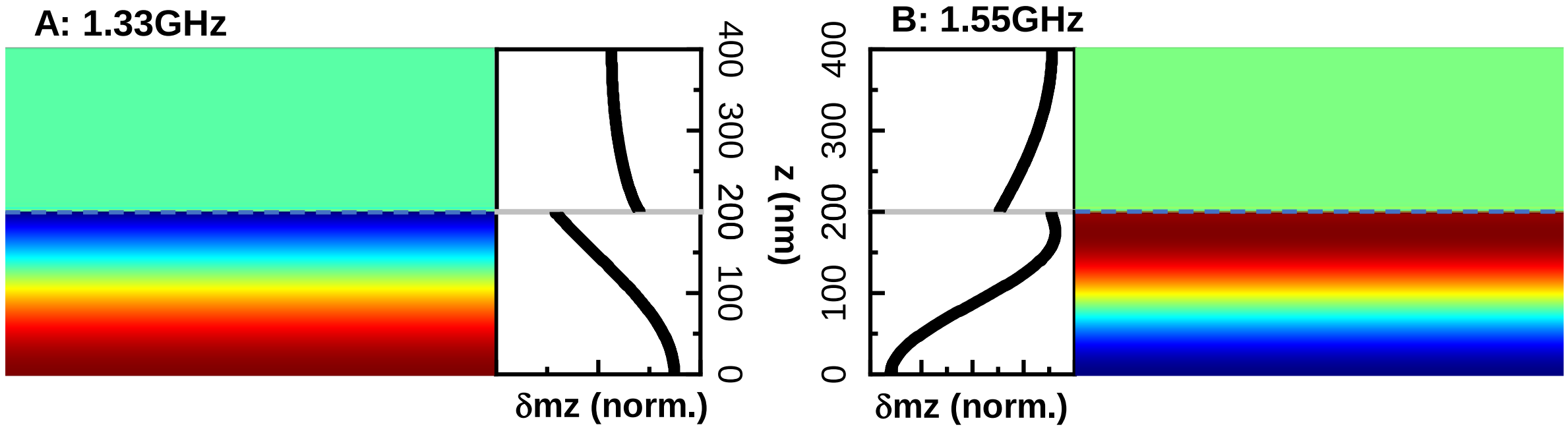

**Supplementary Figure 8 | The two hybrid eigenmodes at 105 Oe.** The $\delta m_z$ distribution of the normalized intensity in the z direction for the two hybrid modes at 1.33 GHz and 1.55 GHz respectively.

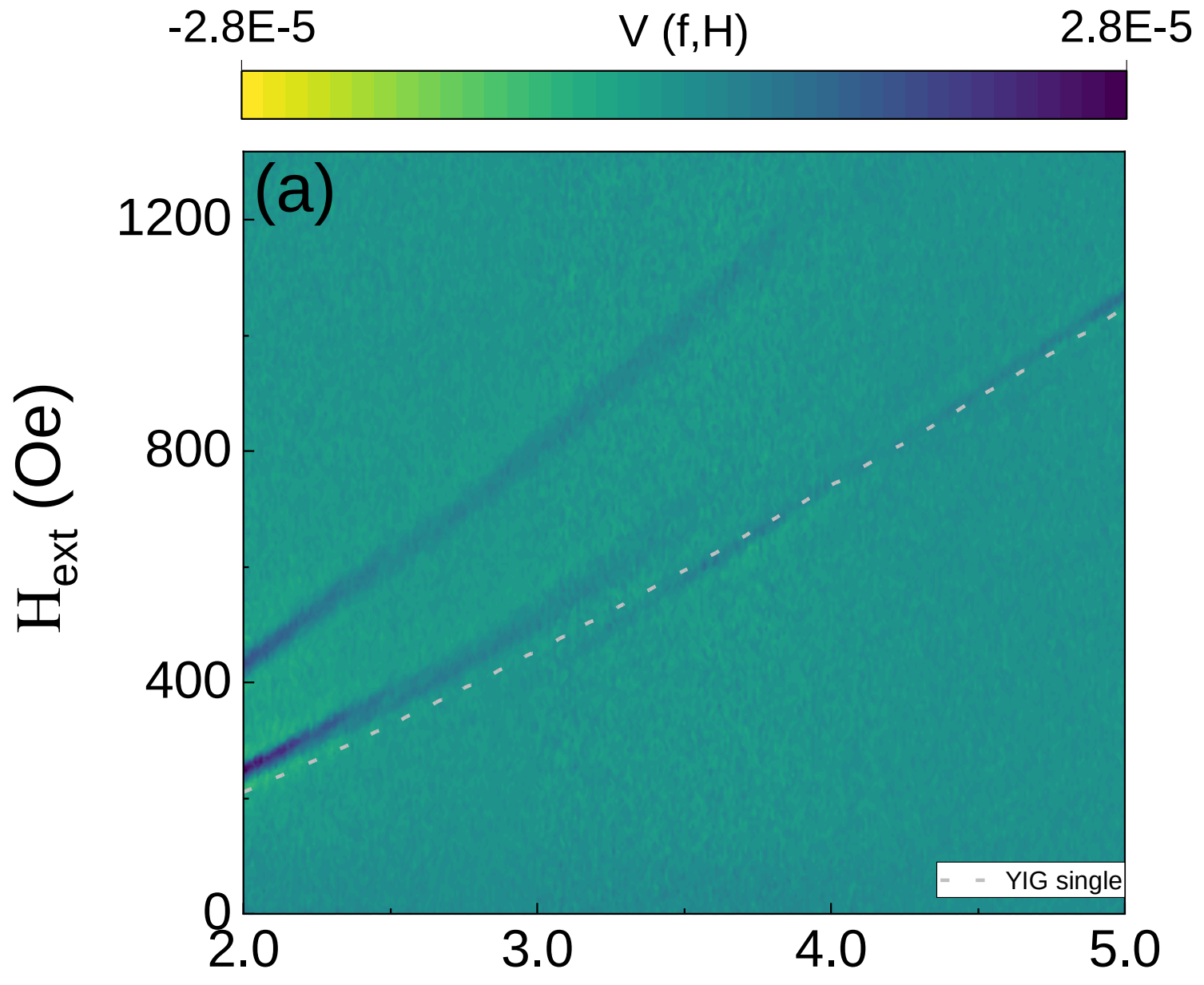

**Supplementary Figure 9 | The spin-wave absorption spectra for TmIG(100 nm)/YIG(100 nm) in comparison with YIG(100 nm)/GGG (grey dashed line).** Lower resonant magnetic field $H_{res}$, or higher resonant frequency $\omega_{res}$, is observed for YIG(100 nm)/GGG both before and after the avoided crossing. This shows that for the TmIG/YIG sample, the interfacial exchange field applied on YIG is opposite to the external magnetic field, which supports the antiferromagnetic coupling nature concluded in the main text.

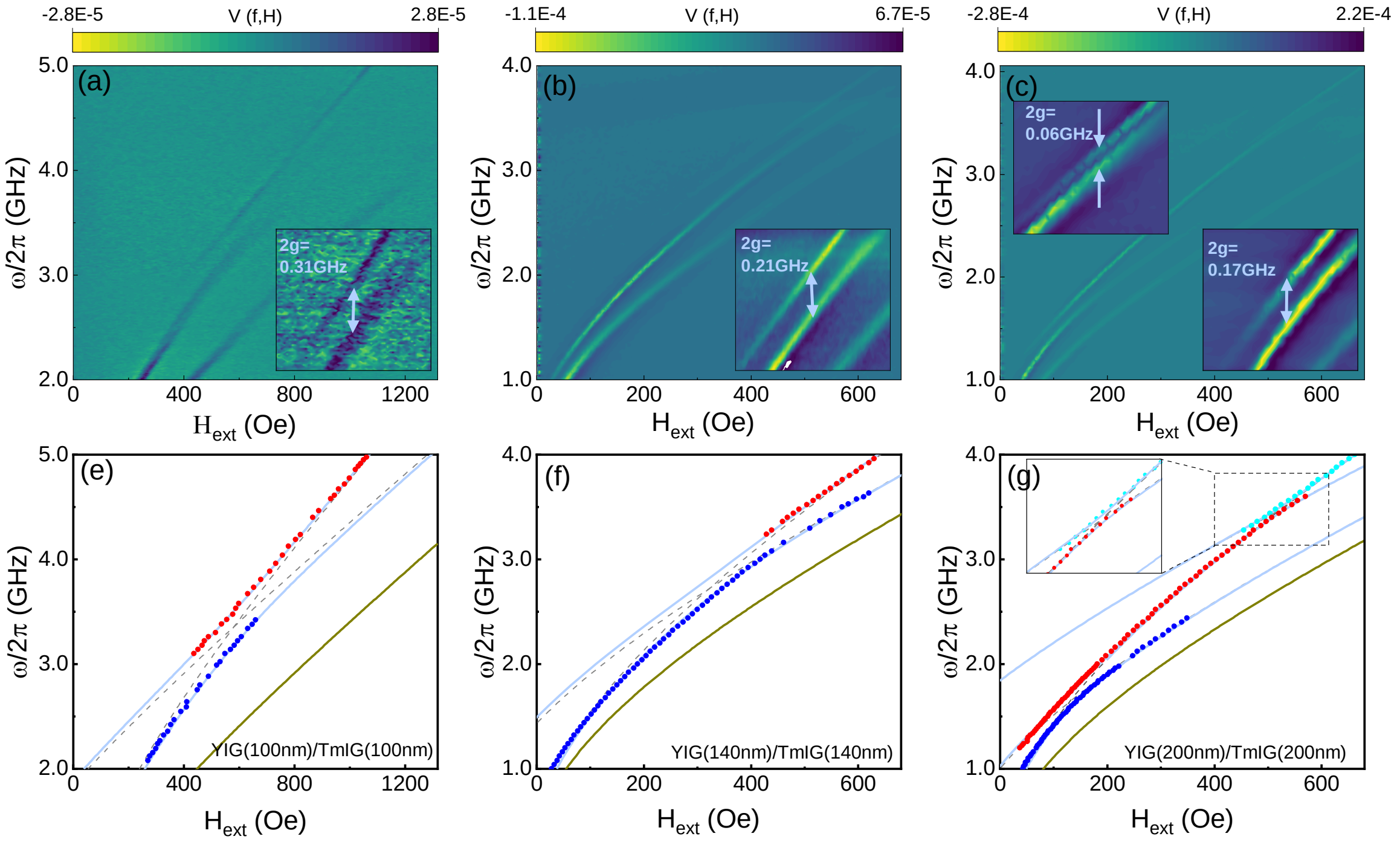

**Supplementary Figure 10 | Magnetic insulator heterostructure thickness dependence of the strong magnon-magnon coupling. a-c,** Experimentally color-coded spin-wave absorption spectra with the YIG (100nm)/TmIG (100nm) (**a**), YIG (140nm)/TmIG (140nm) (**b**), YIG (200nm)/TmIG (200nm) (**c**). Inset magnification was performed at each anti-crossing region. **e-g,** Resonant absorption peaks of the two hybrid modes as a function of external magnetic field with YIG (100nm)/TmIG (100nm) (**e**), YIG (140nm)/TmIG (140nm) (**f**), YIG (200nm)/TmIG (200nm) (**g**) bilayer. Solid curves show the numerical theory method fitting as hybrid modes using Supplementary note2. Data points are extracted from experimental data by fitting the line shapes to two independent derivative Lorentzian functions from **(a-c)**. The increase of effective magnetization in YIG and TmIG and the stiffening of YIG and TmIG resonance frequency is due to the increase of different magnetic insulator thickness.

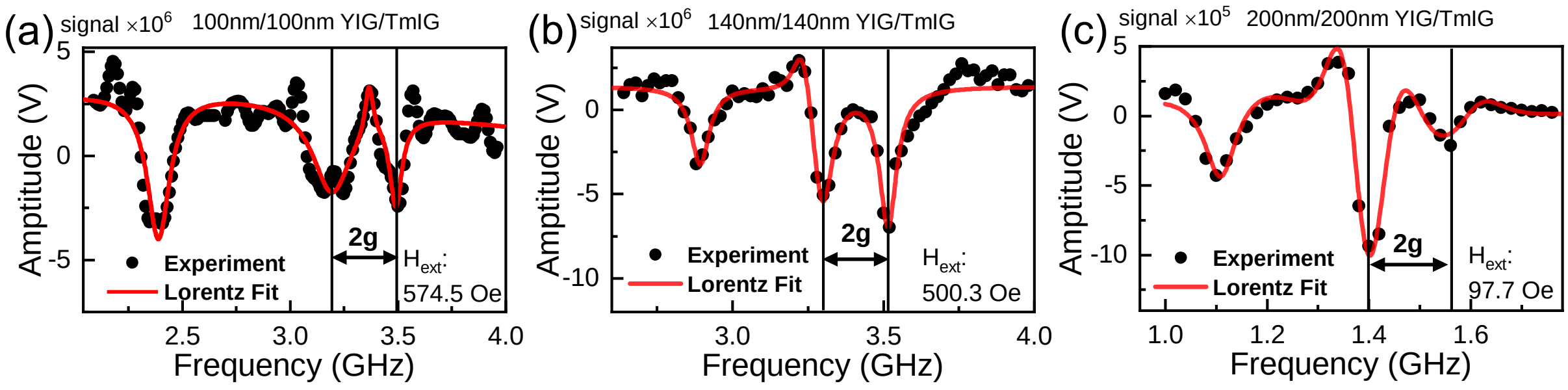

**Supplementary Figure 11 | Magnetic insulator heterostructure thickness dependence of the coupling strength. a-c,** Experimentally spin-wave spectra at the minimum resonance separation with frequency with the TmIG(100 nm)/YIG(100 nm) (**a**), TmIG(140 nm)/YIG(140 nm) (**b**), TmIG(200 nm)/YIG(200 nm) (**c**). At the same time, the red line is fitted by the Lorentz peak function to coupling strength.

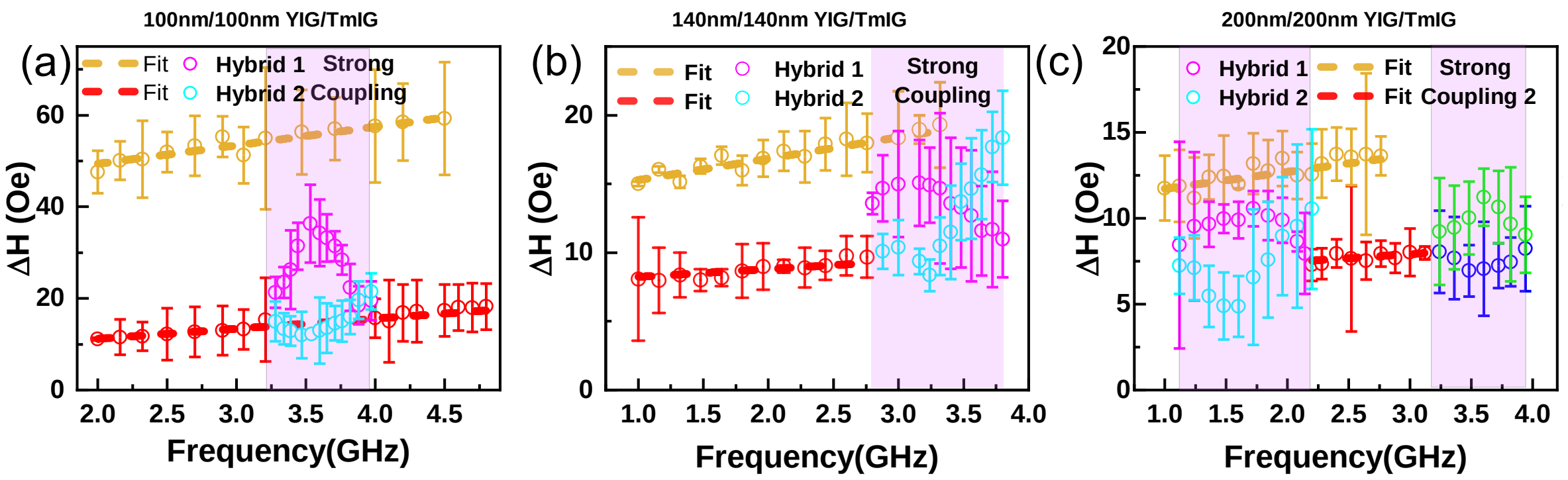

**Supplementary Figure 12 | Thickness dependence of the Gilbert damping. a-c,** Experimentally line-width with frequency with the TmIG(100 nm)/YIG(100 nm) (**a**), TmIG(140 nm)/YIG(140 nm) (**b**), TmIG(200 nm)/YIG(200 nm) (**c**). Circle points with the error bars represent linewidth extracted from experimental data by fitting the line shapes to three independent derivative Lorentzian functions from **(a-c)**. Dashed lines are linear fits away from the strong coupling region through equation 5. The fitting effective damping constants are summarized in the Supplementary Table S2.

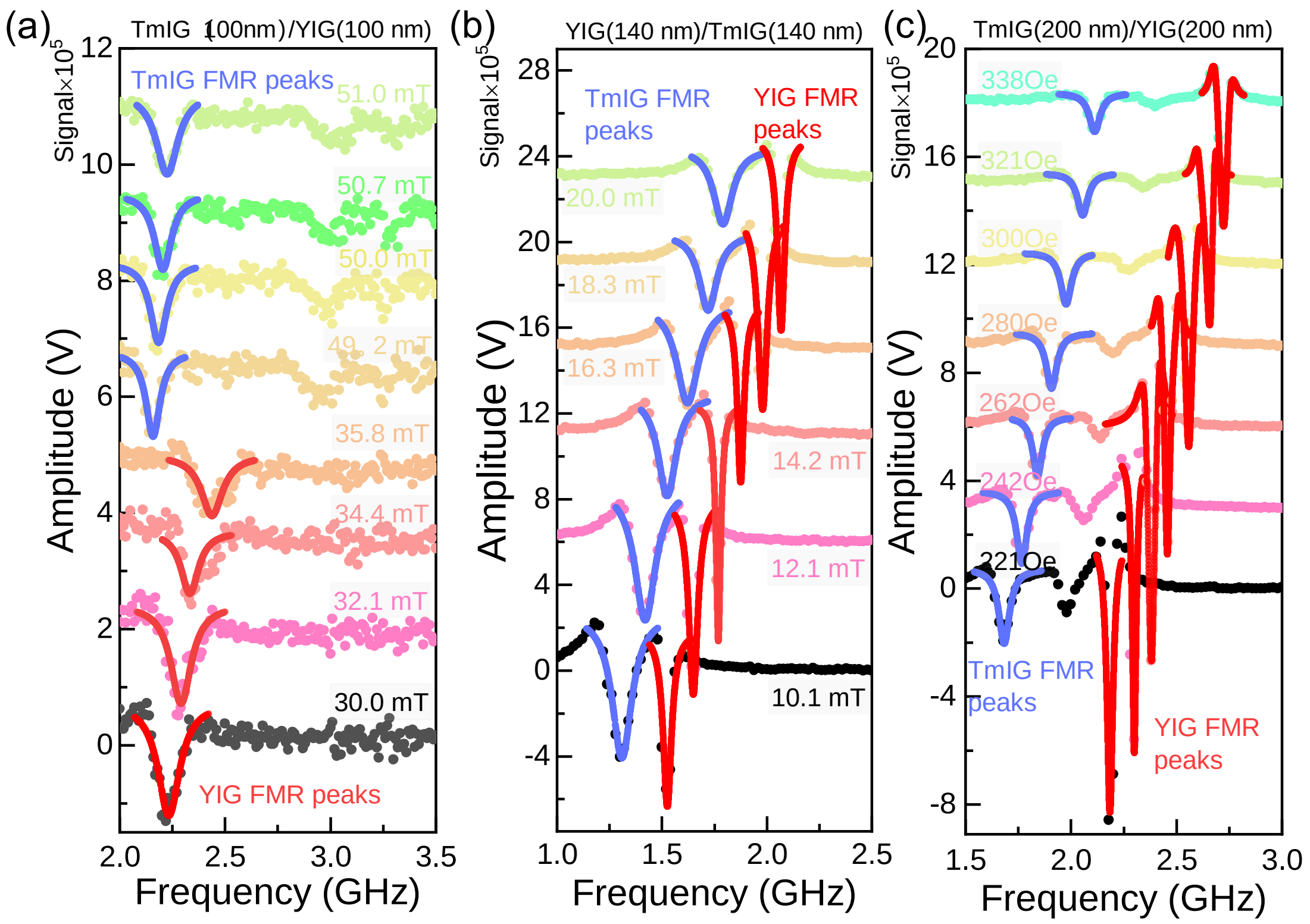

**Supplementary Figure 13 | Thickness dependence of the dissipation rate. a-c,** Frequency sweep curves away from the coupling region. Dissipation rate is obtained by Lorentzian peak function fitting. Red solid line presents YIG FMR, blue solid line presents TmIG FMR.

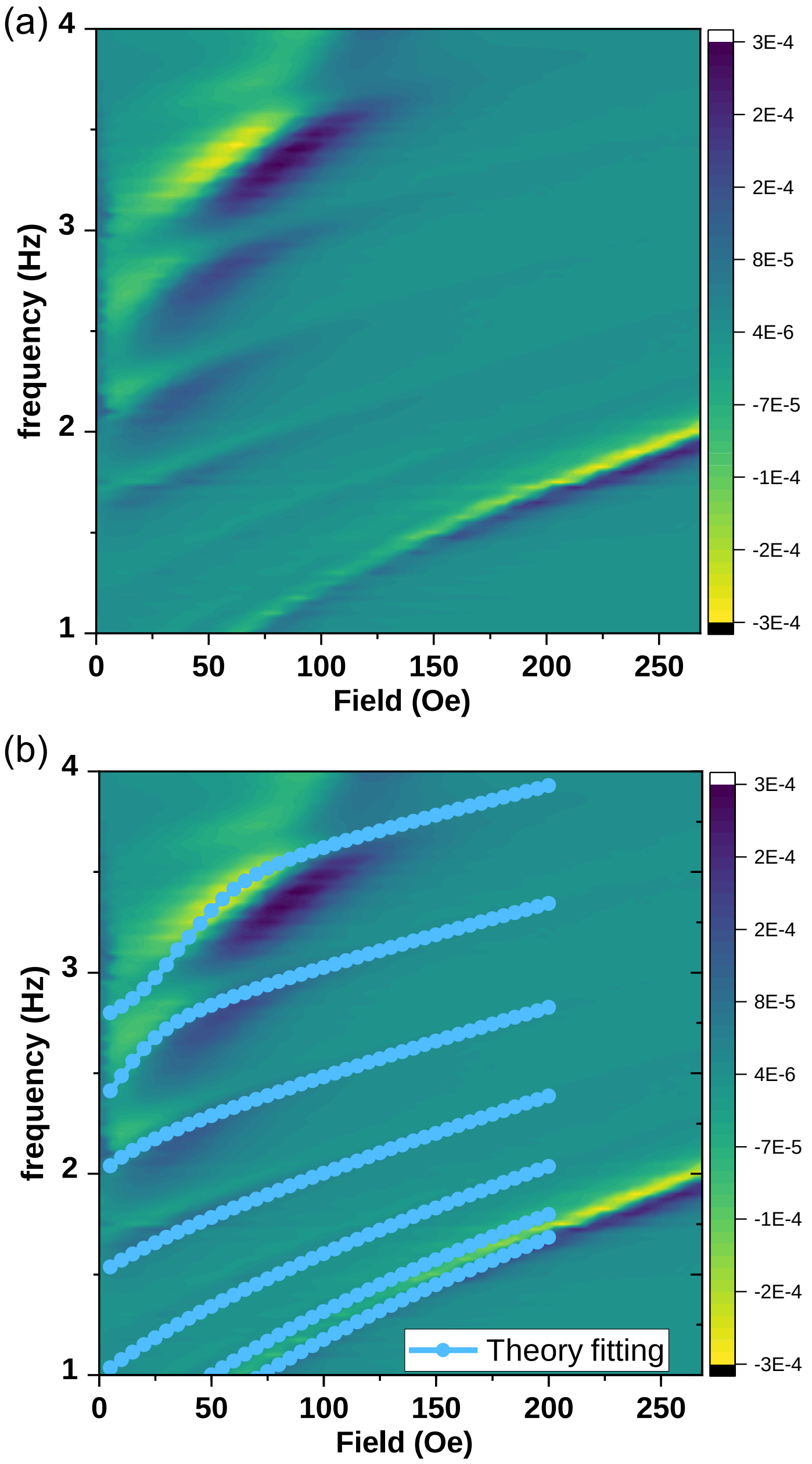

**Supplementary Figure 14 | Magnetic field dependency of resonant peaks for $CoFeB_{50nm}/TmIG_{350nm}$ heterostructure. a,** Experimentally color-coded spin-wave absorption spectra of the $CoFeB_{50nm}/TmIG_{350nm}$ for the first seven resonance modes of TmIG (n=0-6) and the uniform mode of CoFeB (n=0). **b,** Fitting of antiferromagnetic interfacial coupling energy by theoretical inversion.

1. Magnetic heterostructures: advances and perspectives in spinstructures and spintransport. (Springer Verlag, 2008). doi:10.1088/0031-9112/23/4/020.
2. Livesey, K. L., Crew, D. C. & Stamps, R. L. Spin wave valve in an exchange spring bilayer. *Phys. Rev. B* **73**, 184432 (2006).
3. Klingler, S. *et al.* Spin-Torque Excitation of Perpendicular Standing Spin Waves in Coupled YIG / Co Heterostructures. *Phys. Rev. Lett.* **120**, 127201 (2018).
4. Li, Y. *et al.* Coherent Spin Pumping in a Strongly Coupled Magnon-Magnon Hybrid System. *Phys. Rev. Lett.* **124**, 117202 (2020).
5. Zhang, Z., Yang, H., Wang, Z., Cao, Y. & Yan, P. Strong coupling of quantized spin waves in ferromagnetic bilayers. *Phys. Rev. B* **103**, 104420 (2021).
6. Hoffmann, F., Stankoff, A. & Pascard, H. Evidence for an Exchange Coupling at the Interface between Two Ferromagnetic Films. *Journal of Applied Physics* **41**, 1022–1023 (1970).
7. Hoffmann, F. Dynamic Pinning Induced by Nickel Layers on Permalloy Films. *phys. stat. sol. (b)* **41**, 807–813 (1970)
8. Cochran, J. F. & Heinrich, B. Boundary conditions for exchange-coupled magnetic slabs. *Phys. Rev. B* **45**, 13096–13099 (1992).
9. Zhang, J., Yu, W., Chen, X. & Xiao, J. A frequency-domain micromagnetic simulation module based on COMSOL Multiphysics. *AIP Advances* **13**, 055108 (2023).
10. Zhang, X., Zou, C.-L., Jiang, L. & Tang, H. X. Strongly Coupled Magnons and Cavity Microwave Photons. *Phys. Rev. Lett.* **113**, 156401 (2014).
11. Chen, J. *et al.* Strong Interlayer Magnon-Magnon Coupling in Magnetic Metal-Insulator Hybrid Nanostructures. *Phys. Rev. Lett.* **120**, 217202 (2018).
12. Chen, J. *et al.* Excitation of unidirectional exchange spin waves by a nanoscale magnetic grating. *Phys. Rev. B* **100**, 104427 (2019).
13. Khodadadi, B. *et al.* Conductivitylike Gilbert Damping due to Intraband Scattering in Epitaxial Iron. *Phys. Rev. Lett.* **124**, 157201 (2020).
14. Li, Y. *et al.* Giant Anisotropy of Gilbert Damping in Epitaxial CoFe Films. *Phys. Rev. Lett.* **122**, 117203 (2019).
15. Xiong, Y. *et al.* Probing magnon–magnon coupling in exchange coupled $Y_3Fe_5O_{12}$/Permalloy bilayers with magneto-optical effects. *Sci Rep* **10**, 12548 (2020).